\documentclass[%
reprint,
superscriptaddress,
 bibnotes,
 amsmath,
 amssymb,
 aps,
 pra,
]{revtex4-2}
\usepackage{graphicx}
\usepackage[caption=false]{subfig}
\usepackage{qcircuit}
\usepackage{dcolumn}
\usepackage{bm}
\usepackage{tikz}
\usetikzlibrary{shapes}
\usepackage{array}
\newcolumntype{P}[1]{>{\centering\arraybackslash}p{#1}}

\usetikzlibrary{shapes}

\usetikzlibrary{plotmarks}
\usetikzlibrary{decorations.markings}
\usepackage{siunitx}
\usepackage{xy}
\usepackage{dsfont}
\usepackage{pgfplots}
\usepackage{pgf}
\pgfplotsset{compat=1.15}
\usepackage{placeins}
\usepackage{flafter}
\usepackage{color}
\usepackage{extpfeil}
\usepackage{gensymb}
\usepackage{orcidlink}
\usepackage{hyperref}
\hypersetup{
	colorlinks=true,
	linkcolor=blue,
	filecolor=blue,      
	urlcolor=magenta,
	citecolor=magenta,
	pdftex,
	pdfauthor={Ahmed M. Farouk, I.I. Beterov, Peng Xu, S. Bergamini, and I.I. Ryabtsev },
	pdftitle={Parallel implementation of CNOT$^{\text{N}}$ and C$_{2}$NOT$^{2}$ gates via homonuclear and heteronuclear F\"{o}rster interactions of Rydberg atoms},
	pdfcreator={Ahmed M. Farouk},
	}

\usepackage{multirow}
\newcolumntype{M}[1]{>{\centering\arraybackslash}m{#1}}
\usepackage{epstopdf}
\renewcommand{\arraystretch}{1.1}
\makeatletter
\newcommand\tinyv{\@setfontsize\tinyv{5pt}{5}}
\makeatother

\begin{document}

%\preprint{APS/123-QED}

\title{Parallel implementation of CNOT$^{\text{N}}$ and C$_{2}$NOT$^{2}$ gates via homonuclear and heteronuclear F\"{o}rster interactions of Rydberg atoms}% Force line breaks with \\
\author{Ahmed M. Farouk \orcidlink{0000-0002-6230-1234}}
\email[]{ahmed.farouk@azhar.edu.eg}
\affiliation{Novosibirsk State University, 630090 Novosibirsk, Russia}
\affiliation{Rzhanov Institute of Semiconductor Physics SB RAS, 630090 Novosibirsk, Russia}
\affiliation{Faculty of Science, Al-Azhar University, 11884, Cairo, Egypt}
%----%----%----%----%----%----%----%----%----%----%----%----%----%----
\author{I.I. Beterov \orcidlink{0000-0002-6596-6741}}
\email[]{beterov@isp.nsc.ru}
\affiliation{Novosibirsk State University, 630090 Novosibirsk, Russia}
\affiliation{Rzhanov Institute of Semiconductor Physics SB RAS, 630090 Novosibirsk, Russia}
\affiliation{Institute of Laser Physics SB RAS, 630090 Novosibirsk, Russia}
\affiliation{Novosibirsk State Technical University, 630073 Novosibirsk, Russia}

\author{Peng Xu\orcidlink{0000-0001-8477-1643}}
\affiliation{State Key Laboratory of Magnetic Resonance and Atomic and Molecular Physics, Innovation Academy for Precision Measurement Science and Technology, Chinese Academy of Sciences, Wuhan 430071, China}
\affiliation{Wuhan Institute of Quantum Technology, Wuhan 430206, China}

\author{S. Bergamini}
\affiliation{School of Physical Sciences, The Open University, Milton Keynes, MK7 6AA, United Kingdom}

\author{I.I. Ryabtsev}
\affiliation{Novosibirsk State University, 630090 Novosibirsk, Russia}
\affiliation{Rzhanov Institute of Semiconductor Physics SB RAS, 630090 Novosibirsk, Russia}

%----%----%----%----%----%----%----%----%----%----%----%----%----%----
\date{\today}% It is always \today, today,
\begin{abstract}
We analyze schemes of high-fidelity multiqubit CNOT$^{\text{N}}$ and C$_{2}$NOT$^{2}$ gates for alkali-metal neutral atoms used as qubits. These schemes are based on the  electromagnetically induced transparency and Rydberg blockade, as proposed by M. M\"{u}ller et al.  \href{https://doi.org/10.1103/PhysRevLett.102.170502}{[PRL 102, 170502 (2009)]}. In the original paper, the fidelity of multi-qubit CNOT$^{\text{N}}$ gate based on Rydberg blockade was limited by the undesirable interaction between the target atoms, and by the coupling laser intensity.  We propose overcoming these limits by using strong heteronuclear dipole-dipole interactions via F\"{o}rster resonances for control and target atoms, while the target atoms are coupled by weaker van der Waals interaction. We have optimized the gate performance in order to achieve higher fidelity, while keeping coupling laser intensity as small as possible in order to improve the experimental feasibility of the gate schemes. We also considered optimization of schemes of C$_{2}$NOT$^{2}$ gates, where the fidelity is affected by the relation between the control-control, control-target and target-target interaction energies. Our numeric simulations confirm that the fidelity of CNOT$^4$ gate (single control and four target atoms) can be up to $99.3\%$ and the fidelity of C$_2$NOT$^2$ (two control and two target atoms) is up to $99.7\%$ for the conditions which are experimentally feasible.
\end{abstract}

%Use showkeys class option if keyword
\maketitle

%----%----%----%----%----%----%----%----%----%----%----%----%----%----%----%----%----%----%----%----%----%----%----%----%----%----%----%----%----%----%----%----%----%----%----%----%----%----%----%----%----%----%----%----%----%----%----%----%----%----%----%----%----%----%----%----%----%----%----%----%----%----%----

\section{Introduction}

Notable progress in quantum computing in recent years has resulted in first demonstrations of quantum supremacy with superconducting qubits and photons \cite{arute2019quantum, wu2021strong, zhong2021phase}. Ultracold ions and atoms remain promising platforms for a scalable quantum computer \cite{morgado2021quantum}. The  advantage of ultracold atoms is the potential to create quantum registers of thousands of identical qubits on a micrometer scale \cite{graham2022demonstration}. A substantial improvement of two-qubit gate fidelity in quantum registers based on single trapped atoms has been recently demonstrated \cite{zeng2017entangling, levine2019parallel, graham2019rydberg, evered2023high}. Quantum simulations of complex problems of many-body physics and correlated quantum phases of matter can be performed using two-dimensional arrays of hundreds of Rydberg atoms in optical tweezers \cite{scholl2021quantum, ebadi2021quantum}. However, the fidelity of two-qubit gates for neutral atoms still remains limited. These gates use temporary excitation of atoms into Rydberg states \cite{gallagher1988rydberg}. The dipole moments of Rydberg atoms scale as $n^{2}$ \cite{singer2005long}, where $n$ is the principal quantum number. Ground-state atoms do not interact at distances of few microns, but Rydberg atoms do. Therefore the dipole-dipole interaction of Rydberg atoms can be used for implementation of two-qubit gates and creation of entanglement \cite{jaksch2000fast}. The experimentally demonstrated schemes of two-qubit gates are based on the effect of Rydberg blockade: when two atoms are located at short interatomic distance, they cannot be excited simultaneously to Rydberg states by narrow-band laser radiation \cite{lukin2001dipole, urban2009observation, gaetan2009observation, isenhower2010demonstration,levine2019parallel}.

Multi-qubit gates with many target qubits can provide a remarkable speed-up of quantum algorithms. Realization of geometric and swap gates with atomic qubits using antiblockade was discussed in Refs. \cite{su2021dipole, wu2021one, wu2022unselective}. Schemes for multi-control and multi-target gates based on microwave dressing were proposed in Refs. \cite{khazali2020fast, young2021asymmetric, li2022multiple}. A single-step implementation of the  three-qubit controlled gates with Rydberg atoms was reported \cite{sun2021one}. Also, the realization of a two-qubit controlled-PHASE (C$_{\text Z}$ ) gate via single-modulated-pulse off-resonant modulated driving embedded in a two-photon transition for Rb atoms with high-fidelity entanglement to be 0.980(7) was recently reported \cite{fu2022high}.

\begin{figure}[htb!]\centering
	\subfloat[]{\centering
		\includegraphics[width=1.8cm,height=2cm]{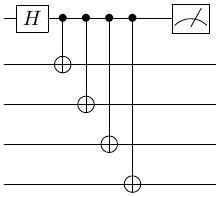}	\label{circuitCNOT}
	}	
	\subfloat[]{\centering
		\includegraphics[width=3.2cm,height=3.4cm]{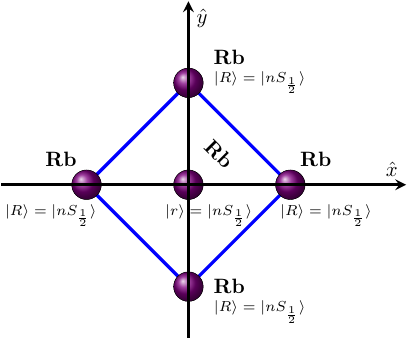}
		\label{homonuclearsymmetric}
	}
	\subfloat[]{\centering
		\includegraphics[width=3.2cm,height=3.4cm]{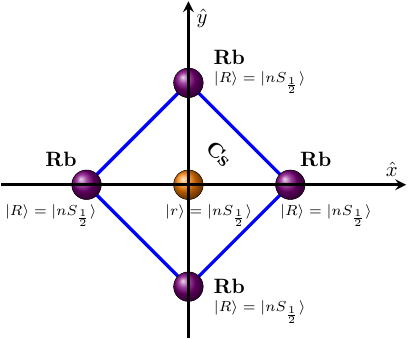}
		\label{heteronuclearAsymmetric}
	}
	\caption{(a) Scheme of generation of multi-atom entangled GHZ-state using a sequence of CNOT gates applied to different target atoms. Firstly, we apply Hadamard gate H on the control atom and do measurement after performing the CNOT gates. (b), (c) Scheme of spatial configurations of Rb and Cs atoms for implementation of the CNOT$^{\text{N}}$ gate in the case of (b) symmetric homonuclear interaction of Rb atoms and (c) asymmetric heteronuclear interaction between Cs control atom and four Rb target  atoms.}
	\label{mainscheme}
\end{figure}

Quantum error correction schemes are of essential importance for quantum information processing with neutral atoms. A scheme for fault-tolerant quantum computing based on surface codes was proposed by Auger et al. \cite{auger2017blueprint}. This scheme requires parallel implementation of multiqubit CNOT$^{\text{N}}$ (where $N$ is a number of target atoms) gates with a single control atom and $N$ target atoms, which are used as ancillary qubits, as shown in Figure~\ref{mainscheme}(a). Recently, a surface code with an atomic quantum processor was experimentally demonstrated \cite{bluvstein2021quantum}. The non-local connectivity between qubits was achieved by coherent transport of qubits in two dimensions and between multiple zones. 

Multi-qubit gates can be built using Rydberg blockade and electromagnetically-induced transparency (EIT) \cite{muller2009mesoscopic}. EIT is a quantum interference phenomenon that can be observed by two optical fields (probe and control lasers) tuned to interact with quantum states of atoms \cite{fleischhauer2005electromagnetically}. The transmission of a weak probe field is enhanced in the presence of a strong (near-)resonant coupling field \cite{ji2021distinction}. Several groups studied EIT with Rydberg states theoretically \cite{pritchard2010cooperative,weimer2010rydberg} and experimentally \cite{peyronel2012quantum,baur2014single, gorniaczyk2016enhancement}. From the original proposal \cite{muller2009mesoscopic} it is clear that  implementation of high-fidelity multi-qubit quantum gates based on EIT requires  large coupling Rabi frequencies (of order of GHz) for transition between low excited and Rydberg states \cite{muller2009mesoscopic}, which is difficult to achieve in the experiment. For lower values of the coupling Rabi frequency, the fidelity of parallel CNOT$^{\text{N}}$ gates becomes substantially limited by the interaction between the target atoms. It is possible to suppress the interaction between the target atoms while keeping strong interaction between control and target atoms by using the dipole-dipole interactions via F\"{o}rster resonances and asymmetric excitation of control and target atoms to different Rydberg states. Moreover, creation of heteronuclear atomic arrays \cite{beterov2015rydberg} provides both extended control over the interatomic interaction and reduced error rates during readout of qubit states. The  heteronuclear atomic species were first entangled in the experiment by Zeng et al. \cite{zeng2017entangling}. A two-dimensional $6 \times 4$ array of two isotopes of Rb atoms was recently demonstrated \cite{sheng2022defect}.  An experimental implementation of a dual-element atomic array with individual control of single Rb and Cs atoms with negligible crosstalk between the two atomic species was recently reported \cite{singh2022dual}. There are also related experimental studies of ultracold dense trapped samples of over 1000 of  $^{87}$Rb $^{133}$Cs molecules in rovibrational ground state with full nuclear hyperfine state control by protocols of stimulated Raman adiabatic passage (STIRAP) with efficiencies of 90\% \cite{takekoshi2014ultracold,molony2014creation}. Rydberg blockade between a single Rb atom and a single RbCs molecule was recently demonstrated \cite{guttridge2023observation}. A mixture of heteronuclear $^6$Li and $^{133}$Cs atomic clouds was studied in Ref. \cite{tung2013ultracold}.

%----%----%----%----%----%----%----%----%----%----%----%----%----%----%----%----%----%----%----%----%----%----%----%----%----%----%----%----%----%----%----%----%----%----%----%----%----%----%----%----%----%----%----%----%----%----%----%----%----%----%----%----%----%----%----%----%----%----%----%----%----%----%----

%----%----%----%----%----%----%----%----%----%----%----%----%----%----%----%----%----%----%----%----%----%----%----%----%----%----%----%----%----%----%----%----%----%----%----%----%----%----%----%----%----%----%----%----%----%----%----%----%----%----%----%----%----%----%----%----%----%----%----%----%----%----%----

In the present work we optimize the performance of a parallel CNOT gate based on Rydberg blockade and EIT in order to reduce the required coupling Rabi frequency for transitions between low excited and Rydberg states and improve the overall gate fidelity. With decrease of the coupling Rabi frequency the target-target interaction deteriorates the gate performance \cite{muller2009mesoscopic}. When both control and target atoms are excited to identical Rydberg states [symmetric homonuclear interaction, as shown in Figure~\ref{mainscheme}(b)], it is not possible to tune the control-target and target-target interactions independently. However, if the control and target atoms are excited to different Rydberg states [asymmetric homonuclear or heteronuclear interaction, which is illustrated in  Figure~\ref{mainscheme}(c)], it is possible to meet the conditions of F\"{o}rster resonance \cite{walker2008consequences, beterov2018adiabatic}  for control-target interaction and to keep target-target interaction in the van der Waals regime. That allows substantial suppression of the target-target interaction and obtaining higher fidelity of parallel CNOT gate at moderate coupling Rabi frequencies of 100-200 MHz, which can be readily achieved in modern experiments for tightly focused laser beams.

While working on this manuscript we became aware of a foremost experimental work \cite{mcdonnell2022demonstration} demonstrating the considered EIT gate protocol for two-qubits, verifying the ability to perform a native CNOT gate. The authors managed to achieve a loss corrected gate fidelity of  $\mathcal{F}_{\text{CNOT}}^{\text{cor}}=0.82(6)$, and prepared an entangled Bell state with $\mathcal{F}_{\text{Bell}}^{\text{cor}}=0.66(5)$ by trapping individually a pair of $^{133}$Cs atoms separated by $6~\mu\text{m}$.

The paper is organized as follows: in section \ref{sectionEIT_gate} we describe the scheme of multi-qubit CNOT$^{\text{N}}$ gate and the physical model used for our numeric simulation of the gate performance. In section \ref{sectionPhysics}, we discuss the properties of the asymmetric homonuclear and heteronuclear F\"{o}rster interactions. In section \ref{sectionFidelity}, we investigate the influence of the  parameters of atomic states and laser fields, F\"{o}rster interaction channels, and gate duration on the gate fidelity in homonuclear symmetric and heteronuclear asymmetric configurations. In section \ref{sectionC2NOT2}, we extend our approach to implement a  C$_2$NOT$^2$ gate with two control and two target atoms  and calculate its fidelity for heteronuclear configuration. Analysis of gate errors due to spontaneous emissions is given in section \ref{sectionGate-error}. In Appendix \ref{Appendix:Multi-Rydberg}, the model of the atomic system is described taking into account multiple Rydberg interaction channels. Results of calculations are compared with the single-channel model. 

%----%----%----%----%----%----%----%----%----%----%----%----%----%----%----%----%----%----%----%----%----%----%----%----%----%----%----%----%----%----%----%----%----%----%----%----%----%----%----%----%----%----%----%----%----%----%----%----%----%----%----%----%----%----%----%----%----%----%----%----%----%----%----

\section{Scheme of Rydberg EIT CNOT$^{\text{N}}$ gate\label{sectionEIT_gate}}

\begin{figure}[htb!]\centering
	\subfloat[Sequence of pulses]{\centering
		\includegraphics[width=6cm,height=4cm]{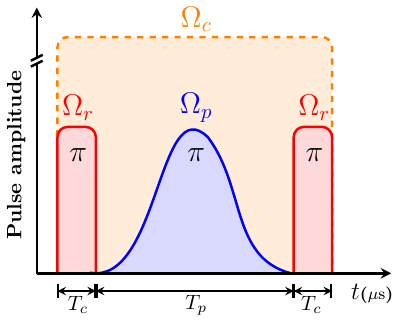}
	}\\
	\subfloat[No transfer scheme]{\centering
		\includegraphics[width=4cm,height=4.5cm]{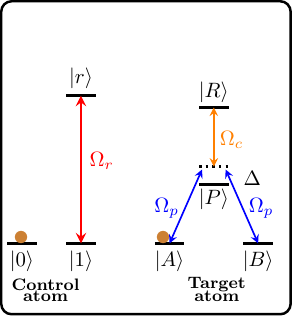}
	}
	\subfloat[Transfer scheme]{\centering
		\includegraphics[width=4cm,height=4.5cm]{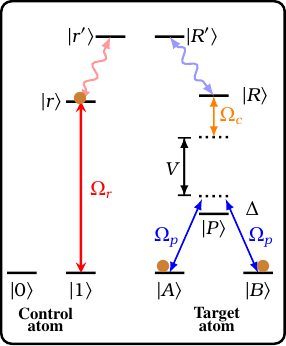}
	}
	\caption{(a) Sequence of laser pulses for Rydberg EIT gate. $\Omega_c$ couples intermediate excited state $|P\rangle$ and Rydberg state $|r\rangle$ of the target atom. Two-photon smooth Raman $\pi$ pulse $\Omega_p(t)$ couples logical states of the target atoms $|A^{N}\rangle$ and $|B^{N}\rangle$ of the target atom. Laser $\pi$ pulses $\Omega_r$ excite and de-excite Rydberg states of control atom;  (b) Scheme of CNOT gate operation in the  regime of blocked population transfer (the control atom in ground state $|0\rangle$). No population transfer between states $|A\rangle$ and $|B\rangle$ is allowed; (c) Scheme of CNOT gate operation in the transfer regime. The ground state of control atom $|1\rangle$ is coupled to the Rydberg state $|r\rangle$ by a $\pi$-pulse $\Omega_r$. The ground states of the target atom $|A\rangle$ and $|B\rangle$ are coupled to the intermediate state $|P\rangle$ by a smooth Raman $\pi$-pulse $\Omega_p(t)$, and the intermediate state is coupled to the Rydberg state $|R\rangle$ by resonant laser radiation with Rabi frequency $\Omega_c$. The dipole-dipole interaction between control and target atoms results from coupling of Rydberg states $|r\rangle$ and $|R\rangle$ to $|r'\rangle$ and $|R'\rangle$, respectively.}
	\label{Fig: scheme}
\end{figure}

The scheme of multiqubit CNOT$^{\text{N}}$ gate, proposed in Ref. \cite{muller2009mesoscopic}, is shown in Figure~\ref{Fig: scheme}(a). The gate operation can be understood as  following: (i) If the control atom is initially in the ground state $|0\rangle$, the first $\pi$-pulse does not change its quantum state. The Raman transfer between states  $|A\rangle$ and $|B\rangle$ is inhibited due to the interaction with intensive resonant coupling radiation with Rabi frequency $\Omega_c$, as shown in Figure~\ref{Fig: scheme}(b).  The fidelity of blocking the population transfer is determined by the value of coupling Rabi frequency $\Omega_c$. (ii) If the control atom is initially prepared in the ground state $|1\rangle$, then it will be excited to the Rydberg state  $|r\rangle$  by the first laser pulse $\Omega_r$. The interaction between control and target atoms will shift the energy levels of the target atoms by a value of  $V_{c \, t_k}$ where $t_k$ denotes k$^{th}$ target atom, and will make the coupling radiation $\Omega_c$ off-resonant for the transition between intermediate excited and Rydberg state of the target atoms, as shown in Figure~\ref{Fig: scheme}(c). Thus, the conditions for EIT are not met anymore, and the Raman population transfer between the states $|A\rangle$ and $|B\rangle$ becomes possible. In the ideal limit of a blockade regime, the Rydberg states of the target atoms are never populated.

%----%----%----%----%----%----%----%----%----%----%----%----%----%----%----%----%----%----%----%----%----%----%----%----%----%----%----%----%----%----%----%----%----%----%----%----%----%----%----%----%----%----%----%----%----%----%----%----%----%----%----%----%----%----%----%----%----%----%----%----%----%----%----

\begin{figure}[htb!]\centering
	\subfloat[]{\centering
		\includegraphics[width=1.8cm,height=1.8cm]{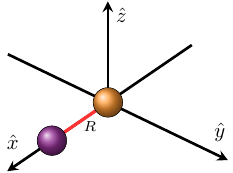}
	}
	%%%%%%%%%%%%%%%%%%%%%%%%%%%%%%%%%%%%%%%%%%%%%%%%%%%%%%%%%%%%%%%%%%%%%%%%%%%%%%%%%%%%%%%%%%%%%%
	\subfloat[]{\centering
		\includegraphics[width=1.8cm,height=1.8cm]{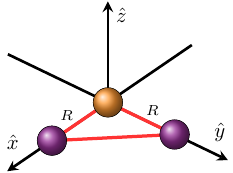}
	}
	%%%%%%%%%%%%%%%%%%%%%%%%%%%%%%%%%%%%%%%%%%%%%%%%%%%%%%%%%%%%%%%%%%%%%%%%%%%%%%%%%%%%%%%%%%%%%%
	\subfloat[]{\centering
		\includegraphics[width=1.8cm,height=1.8cm]{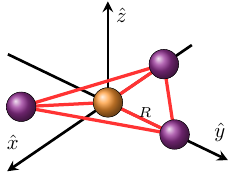}
	}
	%%%%%%%%%%%%%%%%%%%%%%%%%%%%%%%%%%%%%%%%%%%%%%%%%%%%%%%%%%%%%%%%%%%%%%%%%%%%%%%%%%%%%%%%%%%%%%
	\subfloat[]{\centering
		\includegraphics[width=1.8cm,height=1.8cm]{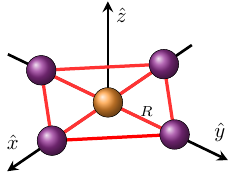}
	}
	\caption{Spatial configurations of control and target atoms. The control atom is at the origin. Here $\hat{z}$ is the quantization axis and $R$ is the interatomic distance between the control and the target atoms for  linear configurations (a) with single target atom, (b) with two target atoms;   (c) Triangular configuration with three target atoms equally displaced from the control atom. The coordinates of the target atoms are $(-R,0,0)$, $(0,R,0)$, and $(R/\sqrt{2},-R/\sqrt{2},0)$; (d) Rectangular configuration with four target atoms. The coordinates of the target atoms are $(R,0,0)$, $(-R,0,0)$, $(0,R,0)$, and $(0,-R,0)$.
		\label{spatialconfiguration}}
\end{figure}

%----%----%----%----%----%----%----%----%----%----%----%----%----%----%----%----%----%----%----%----%----%----%----%----%----%----%----%----%----%----%----%----%----%----%----%----%----%----%----%----%----%----%----%----%----%----%----%----%----%----%----%----%----%----%----%----%----%----%----%----%----%----%----

It is shown in Ref. \cite{muller2009mesoscopic} that the following conditions for CNOT$^\text{N}$ gate are satisfied:
\begin{equation}\label{gate-eq}
	\begin{split}
		\text{(I) No transfer: \hskip0.4cm}&|0\rangle|A^N\rangle \rightarrow 	|0\rangle|A^N\rangle, \\& |0\rangle|B^N\rangle \rightarrow 	|0\rangle|B^N\rangle, \qquad\qquad
	\end{split}
\end{equation}

\begin{equation}\label{gate-eq2}
	\begin{split}
		\text{(II) Transfer: \hskip0.8cm}	& |1\rangle|A^N\rangle \rightarrow -(-1)^{N}	|1\rangle|B^N\rangle, \\&
		|1\rangle|B^N\rangle \rightarrow 	-(-1)^{N} |1\rangle|A^N\rangle.
	\end{split}
\end{equation}

We considered several spatial configurations for $\text{N}=1-4$ target atoms, which are illustrated in Figure~\ref{spatialconfiguration}. The control atom is at the origin. The target atoms are equally displaced from the control atom. For $N=3$ ($N=4$), the target atoms are placed on the vertices' of an isosceles triangle (square) where the distance between the nearest target atoms is $\sqrt{2}R$.

The interaction of the control atom  with radiation in the rotating wave approximation (RWA) is described by the Hamiltonian in basis of $|0\rangle$, $|1\rangle$ and $|r\rangle$ as
\begin{equation}\label{controlhamiltonian}
	\hat{H}_{\text{\tiny C}}=\frac{\hbar}{2}
	\left(\begin{matrix}
		0&0&0\\
		0&0&\Omega_r\\
		0&\Omega_r&0\\
	\end{matrix}\right),
\end{equation}
where $\hbar$ is Planck's constant, and $\Omega_r$ is a sharp $\pi$ pulse which couples the Rydberg state $|r\rangle$ with $|1\rangle$ and is applied for $T_c$~$\mu s$. Explicitly $\Omega_r(t)$ is defined as following

\begin{equation}
	\Omega_r(t)=\left\{
	\begin{array}{ll}
		0,				& t<0.\\
		\dfrac{\pi}{T_{c}},		 &	0\leq t\leq T_{c}.\\
		0,				& T_{c}< t < (T_{p}+T_{c}).\\
		\dfrac{\pi}{T_{c}},		 &	(T_{p}+T_{c}) \leq t\leq(T_{p}+2\,T_{c}).\\
		0,				& t>(T_{p}+2\,T_{c}).
	\end{array}
	\right.
\end{equation}

The  interaction of the target atom with radiation for an inverted Y configuration of the atomic energy levels is described by the Hamiltonian in basis of $|A\rangle$, $|B\rangle$, $|P\rangle$, and $|R\rangle$ as
\begin{equation}
	\hat{H}_{\text{\tiny T}}=\frac{\hbar}{2}
	\left(\begin{matrix}
		0&0&\Omega_p(t)&0\\
		0&0&\Omega_p(t)&0\\
		\Omega_p(t)&\Omega_p(t)&-2\Delta&\Omega_c\\
		0&0&\Omega_c&0\\
	\end{matrix}\right)
	\label{TargetAtomHamiltonian}
\end{equation}
where $\Omega_p(t)=\sqrt{\frac{16\pi\Delta}{3T}}\sin^2(\frac{\pi t}{T})$ is a smooth Raman $\pi$-pulse that couples the ground states of the Rb target atom $|A\rangle=|5\,S_{1/2},F=1 \rangle$ and $|B\rangle=|5\,S_{1/2},F=2 \rangle$ to the intermediate state $|P\rangle=|6\,P_{3/2}, m_j=3/2\rangle$ \footnote{For Cs atoms, the long-lived ground states, and intermediate state of the target atoms are $|A\rangle=|6\,S_{1/2},F=3 \rangle$, $|B\rangle=|6\,S_{1/2},F=4 \rangle$ and $|P\rangle=|7\,P_{3/2}\rangle$, respectively.} with $\int_{0}^{T}dt \, \Omega_p^{2}(t)=2\pi\Delta$. Here $\Delta$ is the detuning from the resonance with the intermediate state $|P\rangle$, as shown in Figure~\ref{Fig: scheme}. Explicitly $\Omega_p(t)$ is defined as following
\begin{equation} 
	\Omega_p(t)=\left\{
	\begin{array}{ll}
		0,																							& t<T_{c},\\
		\sqrt{\frac{16\pi\Delta}{3T_{p}}}\sin^2\left(\frac{\pi}{T_{p}}(t-T_{c})\right),		 &	T_{c} \leq t\leq(T_{p}+T_{c}),\\
		0,																							& t > (T_{p}+T_{c}).\\
	\end{array}
	\right.
\end{equation}
The value $\Omega_c$ is a constant Rabi frequency which couples the intermediate state $|P\rangle$ with Rydberg state $|R\rangle=|n\, S_{1/2}\rangle$ [see Figure~\ref{Fig: scheme}(b)]. 

The model Hamiltonian of the combined system with single control atom and \textit{N} target atoms can be written as
\begin{equation}
	\begin{split}
		\hat{H}=\hat{H}_{\text{\tiny C}}\otimes \hat{\mathds{1}}_{T} &+ \hat{\mathds{1}}_{\text{\tiny C}} \otimes \hat{\mathcal{H}}_{\text{\tiny T}} +
		%%%%%%%%%%%%%%%%%%%%%%%%%%%
		%%%%Control-Target interaction	
		\hat{\mathcal{H}}_{\text{\tiny CT}}
		%%%%%%%%%%%%%%%%%%%%%%%%%%%
		%%%%Target-Target interaction
		+\hat{\mathcal{H}}_{\text{\tiny TT}}
	\end{split}\label{GeneralHamiltonian1}
\end{equation}

where $\hat{\mathds{1}}_{\text{\tiny C}}=I_{4}$ and $\hat{\mathds{1}}_{\text{\tiny T}}=\otimes^{N}I_{5}$ are the identity matrices acting on the control atom, and on the ensemble of target atoms, respectively. $\hat{\mathcal{H}}_{\text{\tiny T}}$ is the Hamiltonian describing the ensemble of target atoms. $\hat{\mathcal{H}}_{\text{\tiny T}}$ can be written in the following form
\begin{equation}
	\hat{\mathcal{H}}_{\text{\tiny T}}=\sum_{i}^{N}\otimes_{j}^{N} \mathds{L}_{ij}
\end{equation}
where 
\begin{equation}
	\mathds{L}_{ij}=\left\{
	\begin{array}{cc}
		\hat{H}_{\text{\tiny T}}, &\text{if } i=j.\\
		\hat{\mathds{1}}_{\text{\tiny T}}, & \text{otherwise.}\\
	\end{array}
	\right.
\end{equation}
The third and fourth terms of Eq. (\ref{GeneralHamiltonian1}) are the terms describing the interaction between control and target atoms, and the interaction between the target atoms, respectively. In section \ref{sectionPhysics}, we show how to calculate the interaction energies in homonuclear and heteronuclear architectures.

%$V_{\text{\tiny C\,T}_j}$ ($V_{\text{\tiny T}_j \, \text{\tiny T}_k}$) denotes the interaction energy as a function of interatomic distance between control atom and target atom $j$ (target atom $j$ and target atom $k$). 
%%%%%%%%%%%%%%%%%%%%%%%%%%%%%%%%%%%%%%%%%%%%%%%%%%%%%%%%%%%%%%%%%%%%%%%%%%%%%%%%%%%%%%%%%%%%%%%%%%%%%%%%%%%%%%%%%%%%%%%%

\begin{figure}[htb!]\centering
	\subfloat[$\Omega_p=2\pi\times50$~\si{\MHz}]{\centering
		\includegraphics[width=4cm,height=4cm]{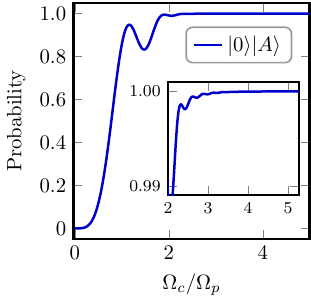}
	}
	\subfloat[$\Omega_c=0.15\,\Omega_p$]{\centering
		\includegraphics[width=4cm,height=4cm]{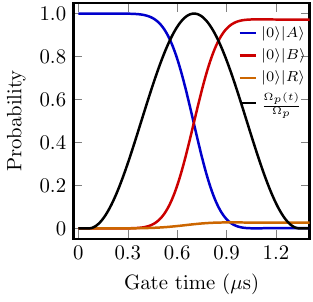}
	}\\
	\subfloat[$\Omega_c=2.0\,\Omega_p$]{\centering
		\includegraphics[width=4cm,height=4cm]{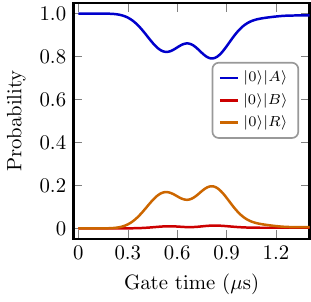}
	}
	\subfloat[$\Omega_c=8.0\,\Omega_p$]{\centering
		\includegraphics[width=4cm,height=4cm]{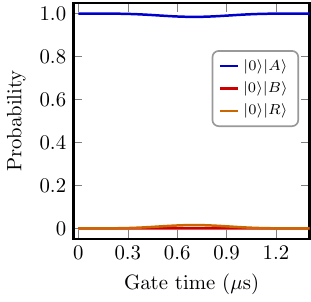}
	}
	\caption{(a) The dependence of the fidelity of blocking the population transfer $|0\rangle |A\rangle \rightarrow |0\rangle |A\rangle$ on the ratio between $\Omega_c$ and $\Omega_p$.  The inset shows the behavior of the population of initial state in the region when it is close to 1. (b-d)  Time dependence of the population of the collective states $P_{|0\rangle|A\rangle}$ (solid-blue curve) and $P_{|0\rangle|B\rangle}$ (solid-red curve) for the case of blocked population transfer during CNOT gate for (b)  $\Omega_c=2\pi\times7.5$~\si{MHz}; (c) $\Omega_c=2\pi\times 100$~\si{MHz} and (d) $\Omega_c= 2 \pi \times 400$~\si{MHz}. The solid-orange curve illustrates the population transfer to the Rydberg state of the target atom in the case of blocked population transfer between logical states.}
	\label{fig:probabilitiesN1}
\end{figure}

%%%%%%%%%%%%%%%%%%%%%%%%%%%%%%%%%%%%%%%%%%%%%%%%%%%%%%%%%%%%%%%%%%%%%%%%%%%%%%%%%%%%%%%%%%%%%%%%%%%%%%%%%%%%%%%%%%%%%%%%

In our simulations, we choose the maximum value of the Raman pulse $\Omega_p=2\pi\times50$~\si{MHz} and the detuning $\Delta=2\pi\times 1200$~\si{MHz}. The Raman pulse is applied for a duration $T_{p}=\frac{16\pi\Delta}{3\text{ max}(\Omega_p)^2}$. Also, we adopted an approach of a non-Hermitian Hamiltonian to consider the finite lifetime of Rydberg state $|r\rangle$ of control atom by adding the term $-\frac{i}{2}\gamma_r |r\rangle \langle r|$ to Eq.~(\ref{controlhamiltonian}) and the finite lifetime of the intermediate states $|P\rangle$ of target atoms by adding the term $-\frac{i}{2}\gamma_p |P\rangle \langle P|$ to Eq.~(\ref{TargetAtomHamiltonian}) where $\gamma_r$ and $\gamma_p$ are the decay rates of Rydberg and intermediate excited states, respectively.

Following the original work \cite{muller2009mesoscopic}, in Figure~\ref{fig:probabilitiesN1} we illustrate the dependence of the fidelity of blocking the population transfer  $|0\rangle |A\rangle \rightarrow |0\rangle |A\rangle$ as a function of the ratio between $\Omega_c$ and $\Omega_p$. In this case, the control atom in not excited to Rydberg state. Therefore, there is no interaction between control and target atoms. As clearly seen in Figure~\ref{fig:probabilitiesN1}(a), for  $\Omega_c>2\,\Omega_p$ in the EIT regime \cite{fleischhauer2005electromagnetically} the transfer between ground states of the target atom is blocked. The time dependence of the population transfer is shown in Figs.~\ref{fig:probabilitiesN1}(b)-(d). At low value of $\Omega_c=0.15\,\Omega_p$ the population transfer  $|A \rangle \to |B \rangle$ becomes allowed [Figure~\ref{fig:probabilitiesN1}(b)]. In the intermediate case of relatively small $\Omega_c=2.0\, \Omega_p$, the state $| A \rangle$ is temporarily depopulated, but finally the system returns to the initial state [see Figure~\ref{fig:probabilitiesN1}(c)]. At very high values $\Omega_c=8.0\, \Omega_p$ the system mostly remains in the state $| A \rangle$  [see Figure~\ref{fig:probabilitiesN1}(d)]. Although the regime of strong coupling is advantageous for maximum gate fidelity, it requires high coupling Rabi frequencies of order of hundreds of MHz or GHz which are difficult to achieve experimentally for highly excited Rydberg states due to the drop of transition matrix elements for ground-Rydberg laser excitation as $n^{-3/2}$. Therefore, the intermediate values of $\Omega_c$ are of interest for experimental implementation. Similar behavior of fidelity is observed for larger number of target atoms.

M\"{u}ller et al. \cite{muller2009mesoscopic} analyzed the effect of target-target interactions in the three-qubit GHZ state for three target atoms while considering the control-target interaction to be constant and varying the target-target interaction. This two-dimensional configuration is not relevant to the typical experimental conditions, since changing the values of the target-target interaction requires varying the interatomic distance between target atoms, which also results in change of the distance between the control and target atoms in all possible spatial configurations of atomic ensembles. In our following simulations, we varied the interatomic distances for several spatial configurations, which are illustrated in Figure~\ref{spatialconfiguration}. In Appendix \ref{Appendix:Multi-Rydberg}, we developed an analogous model for the one described in this section, where the coupling between many Rydberg states in considered. 

%----%----%----%----%----%----%----%----%----%----%----%----%----%----%----%----%----%----%----%----%----%----%----%----%----%----%----%----%----%----%----%----%----%----%----%----%----%----%----%----%----%----%----%----%----%----%----%----%----%----%----%----%----%----%----%----%----%----%----%----%----%----%----

\section{Homonuclear and heteronuclear interaction energy\label{sectionPhysics}}

Our approach is based on the additional control of the energy of interatomic dipole-dipole interactions using F\"{o}rster resonances between two distinguishable atoms, which were studied in Ref. \cite{beterov2015rydberg}. A F\"{o}rster resonance \cite{forster1965modern} means that the energies of two collective states of two Rydberg atoms coupled by dipole-dipole interaction are equal. This enhances the probability of the population transfer between the collective two-body states which is equivalent to energy transfer between interacting atoms \cite{saffman2005analysis, ryabtsev2010observation, tretyakov2017observation}. F\"{o}rster resonance is employed to compromise the choice of selected Rydberg states. The interaction between atoms lies in two different regimes: dipole-dipole regime (d-d) where interaction energy can be described as $V(R)=\frac{C_{3}}{R^3}$ or van der Waals regime (vdW) where $V(R)=\frac{C_{6}}{R^6}$ \cite{browaeys2020many}.  Recently, software packages \cite{vsibalic2017arc, Weber2017} facilitated the calculations of interaction energies between alkali and alkaline earth atoms. We used Alkali Rydberg Calculator (ARC) \cite{vsibalic2017arc} to calculate the  interaction energy of two alkali atoms  in $|n, \ell, j, m_j\rangle$ Rydberg states (here $n$ is the principal quantum number,  $\ell$ is orbital angular momentum number, $j$ is the total angular momentum number, and $m_j$ is the projection of the total angular momentum on the quantization $z$-axis) for homonuclear and heteronuclear configurations. We also calculated the Le Roy radius $R_{\text{\tiny LR}}$\cite{le1974long} which is the internuclear distance between two interacting atoms at which the theory of Le Roy-Bernstein is satisfied, and the interaction potential can be approximated by charge independent atomic distributions. This radius sets the minimum limit of the interatomic distance where our calculations of the interaction energies are valid. As shown in Figure~\ref{spatialconfiguration}, we considered the quantization $z$-axis perpendicular to the interatomic axis, which is the general case for all spatial configurations which are studied in this work. As shown in section \ref{sectionEIT_gate}, we investigated the performance of CNOT gate based on EIT with a single control atom for different number of target atoms for  homonuclear and heteronuclear configurations. The target atoms are identical in all cases.

\subsection{Heteronuclear architecture}
We consider a Cs atom excited from the ground state $|1\rangle$ to Rydberg state $|r\rangle=|81S_{1/2}, m_j=-1/2 \rangle$ as control qubit (as described in section \ref{sectionEIT_gate}), and  spatially ordered Rb atoms as target qubits. The intermediate state $|P\rangle=|6P_{3/2}\rangle$ of target atoms is coupled  to Rydberg state $|R\rangle=|77S_{1/2}, m_j=1/2\rangle$. Due to the interaction between these two atoms the Rydberg states in both atoms will be coupled to other neighboring  Rydberg state(s) $|r;R \rangle \rightarrow |r';R' \rangle$ [see Figure \ref{Fig: scheme}(c)]. The most dominant interaction channel, as shown in Appendix \ref{Appendix:Multi-Rydberg}, is 
\begin{equation}
	\begin{split}
	|81S_{1/2}, m_j&=-1/2; 77S_{1/2}, m_j=1/2\rangle \rightarrow \\& |80P_{1/2}, m_j=1/2; 77P_{3/2}, m_j=3/2\rangle
	\end{split}
\end{equation}
The asymmetric interaction between these two atoms lies in the regime of dipole-dipole interaction. The interaction Hamiltonian between control and target atoms $\hat{\mathcal{H}}_{\text{\tiny CT}}$ can be written as \cite{yu2022multiqubit}
\begin{equation}
	\begin{split}
		\hat{\mathcal{H}}_{\text{\tiny CT}}=\sum_{\kappa}^{\mathcal{N}_{\text{\tiny T}}}\sum_{j}^{N} &\frac{C_3^{(j)}(1-3\cos^2\theta_{\text{\tiny CT$_j$}})}{R_{\text{\tiny CT$_j$}}^{3}}|r\rangle\otimes_{i=1}^{N} \left| \mathds{W}_{ij}\right\rangle_{\kappa} \otimes\\&\otimes \langle r'| \otimes_{i=1}^{N} \prescript{}{\kappa}{\left\langle \mathds{W'}_{ij}\right|} 
		+\sum_{\kappa}^{\mathcal{N}_{\text{\tiny T}}} \sum_{j}^{N} \delta_{F}^{(j)} |r\rangle \otimes \\& \otimes_{i=1}^{N} \left| \mathds{W}_{ij}\right\rangle_{\kappa} \langle r| \otimes_{i=1}^{N} \prescript{}{\kappa}{\left\langle \mathds{W}_{ij}\right|}
	\end{split}\label{Control-TargetinteractionHamiltonian}
\end{equation}
where $C_3=2\pi \times 10$~\si{GHz}.$\mu\text{m}^3$ is a dipole-dipole interaction coefficient for the target atom $j$ separated from the control atom by distance $R_{\text{\tiny CT$_j$}} > R_{\text{\tiny LR}}=2~\mu\text{m}$, $\theta_{\text{\tiny CT$_j$}}=\pi/2$ is the angle between the quantization axis and the interatomic axis. The energy defect $\delta_{F}=2\pi\times2$~\si{MHz} is the energy difference between the collective two-atom Rydberg states for the dominant interaction channel. The collective state $ \otimes_{i=1}^{N} |\mathds{W}_{ij}\rangle = |\mathds{W}_{1j}\rangle | \mathds{W}_{2j} \rangle \dots | \mathds{W}_{Nj}\rangle = | \mathds{W}_{1j} ; \mathds{W}_{2j} ; \dots \mathds{W}_{Nj} \rangle$ is defined as 
\begin{equation}
	\left| \mathds{W}_{ij}\right\rangle_{\kappa}=\left\{
	\begin{array}{cc}
		\left|R\right\rangle, & \text{if } i=j.\\
		\left|\psi_{\text{\tiny T}}\right\rangle_{\kappa}, & \text{otherwise}.
	\end{array}
	\right.
\end{equation}
where $\left|R\right\rangle$ is the excited Rydberg state of the target atom, and $\left|\psi_{\text{\tiny T}}\right\rangle=\{|A\rangle,|B\rangle,|P\rangle, |R\rangle,|R'\rangle\}$ is the set of all basis states of any target atom with dimensions $\mathcal{N}_{\text{\tiny T}}=5$. 

In this architecture, all target atoms are identical, and they interact in the vdW regime. The Hamiltonian describing their interaction can be written as
\begin{equation}
	\begin{split}
		\hat{\mathcal{H}}_{\text{\tiny TT}}=\sum_{\ell=1}^{\mathcal{N}_{\text{\tiny C}}} \sum_{j=1}^{\mathcal{N}_{\text{\tiny T}}} \sum_{l =1}^{N-1} \sum_{k>l}^{N}\frac{C_{6}^{(lk)}}{R_{\text{\tiny T$_{l}$T$_{k}$}}^{6}} \left| \psi_{\text{\tiny C}}\right\rangle_{\ell} &\otimes_{i}^{N} | \mathds{G}_{i}^{(lk)} \rangle \otimes \\&
		\prescript{}{\ell}{\left\langle \psi_{\text{\tiny C}}\right|} \otimes_{i}^{N} \langle \mathds{G}_{i}^{(lk)}|
	\end{split}\label{Target-TargetinteractionHamiltonian}
\end{equation}
where $C_6=2\pi\times2036$~\si{GHz}.$\mu\text{m}^6$ is van der Waals coefficient calculated by fitting the model function with the calculated energy level using ARC function \textit{getC6fromLevelDiagram} for \textit{rStart}=$R_{\text{\tiny LR}}$, \textit{rStop}=$20$~$\mu\text{m}$, and \textit{minStateContribution}=0. The interatomic distance between different target atoms $R_{\text{\tiny T$_{l}$T$_{k}$}}>R_{\text{\tiny vdW}}=4.5~\mu\text{m}$. The set of all basis states of the control atom $\left| \psi_{\text{\tiny C}}\right\rangle=\{ |0\rangle, |1\rangle, |r\rangle, |r'\rangle\}$ with $\mathcal{N}_{\text{\tiny C}}=4$. The collective state $\otimes_{i=1}^{N} |\mathds{G}_{i}^{(lk)} \rangle=$~$| \mathds{G}_{1}\rangle \dots | \mathds{G}_{l} \rangle \dots | \mathds{G}_{k} \rangle \dots | \mathds{G}_{N} \rangle = | \mathds{G}_{1}; \dots \mathds{G}_{l}; \dots \mathds{G}_{k}, \dots, \mathds{G}_{N}\rangle$ where

\begin{equation}
	|\mathds{G}_{i}\rangle=\left\{
	\begin{array}{cc}
		|R\rangle_{l(k)}, & \text{if } i=l \, \mid \, i=k.\\
		|\psi_{\text{\tiny T}}\rangle_{j}, & \text{otherwise}.
	\end{array}
	\right.
\end{equation}

%----%----%----%----%----%----%----%----%----%----%----%----%----%----%----%----%----%----%----%----%----%----%----%----%----%----%----%----%----%----%----%----%----%----%----%----%----%----%----%----%----%----%----%----%----%----%----%----%----%----%----%----%----%----%----%----%----%----%----%----%----%----%----

\subsection{Homonuclear architecture}
The case when all interacting atoms are the same atomic species (Rb), all interactions are in the vdW regime, since all atoms all excited to the same Rydberg state $|77S_{1/2},m_j=1/2\rangle$ and the interaction is in the vdW regime. The control-target Hamiltonian in this case will be in the following form
\begin{equation}
	\hat{\mathcal{H}}_{\text{\tiny CT}} = \sum_{\kappa}^{\mathcal{N}_{\text{\tiny T}}} \sum_{j}^{N} \frac{C_6^{(j)}}{R_{\text{\tiny CT$_j$}}^{6}} |r\rangle\otimes_{i=1}^{N} \left| \mathds{W}_{ij}\right\rangle_{\kappa} \langle r| \otimes_{i=1}^{N} \prescript{}{\kappa}{\left\langle \mathds{W}_{ij}\right|}
\end{equation}
and the target-target interaction Hamiltonian will be the same as given in equation \ref{Target-TargetinteractionHamiltonian} for  Rb atoms. For Cs homonuclear interaction with all atoms are excited to the same Rydberg state $|81S_{1/2},m_j=-1/2\rangle$, the van der Waals coefficient $C_6=2\pi \times 2364$~\si{GHz}.$\mu\text{m}^6$.
%----%----%----%----%----%----%----%----%----%----%----%----%----%----%----%----%----%----%----%----%----%----%----%----%----%----%----%----%----%----%----%----%----%----%----%----%----%----%----%----%----%----%----%----%----%----%----%----%----%----%----%----%----%----%----%----%----%----%----%----%----%----%----

\begin{figure}[htb!]\centering
	\subfloat[]{
		\includegraphics[width=5cm,height=5.6cm]{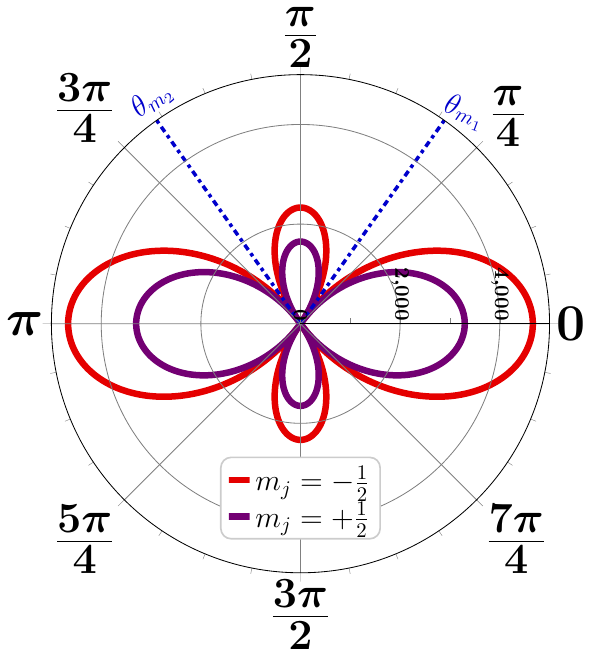}
	}\\
	\subfloat[]{
		\includegraphics[width=4cm,height=5cm]{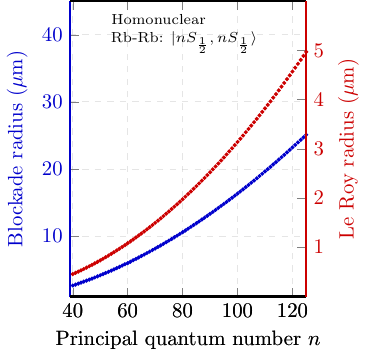}
	}
	\subfloat[]{
		\includegraphics[width=4cm,height=5cm]{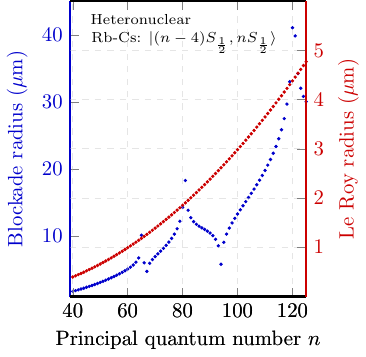}
	}
	\caption{(a) The  dipole-dipole interaction coefficient  $C_3$~(\si{MHz}.$\mu\text{m}^{3}$) as a function of the angle $\theta$ between the interatomic axis and the quantization axis for the heteronuclear interaction between Rb  $|77S_{1/2},1/2 \rangle$ and Cs $|81S_{1/2}, m_{j} \rangle$. Red curve (Violet curve) represent the projection of the total angular momentum on $z$-axis $m_{j}=-\frac{1}{2}$ ($m_{j}=\frac{1}{2}$) for the most dominant interaction channel. (b) and (c) The evolution of blockade radius (blue curve) and Le Roy radius (red curve) as a function of the principal quantum number $n$ for homonuclear interaction of two Rb atoms and heteronuclear interaction between Rb and Cs atoms, respectively.} \label{Fig: hyperfinescheme}
\end{figure}

%----%----%----%----%----%----%----%----%----%----%----%----%----%----%----%----%----%----%----%----%----%----%----%----%----%----%----%----%----%----%----%----%----%----%----%----%----%----%----%----%----%----%----%----%----%----%----%----%----%----%----%----%----%----%----%----%----%----%----%----%----%----%----

In Figure~\ref{Fig: hyperfinescheme}(a), we show the dependence of the dipole-dipole interaction coefficient $C_3$ for Cs and Rb atoms excited to Rydberg states, on the angle $\theta$ between quantization $z$-axis and the interatomic axis for two different cases of the projection of the total angular momentum on the quantization $z$-axis. We have $\theta=\pi/2$ which does not meet the maximum value of interaction energies but for using negative projection of the Rydberg state of Cs atom strengths the interaction. The interaction between the two atoms vanish at the magic angle $\theta_{m_1}=54.7356\degree$, and $\theta_{m_2}=\pi-\theta_{m_1}$ . 

In Figure~\ref{Fig: hyperfinescheme}(b) and \ref{Fig: hyperfinescheme}(c), we show the evolution of blockade radius (dotted-blue curve) and Le Roy radius (dotted-red curve) as a function of the principal quantum number $n$ of the excited Rydberg state for homonuclear interaction between two Rb atoms excited symmetrically to Rydberg states $|nS_{\frac{1}{2}},m_j=1/2$ and the heteronuclear interaction between Rb and Cs atom excited asymmetrically to Rydberg states $|(n-4)S_{\frac{1}{2}},m_j=1/2\rangle$, and $|nS_{\frac{1}{2},m_j=-1/2}$, respectively. It is noted that for homonuclear interactions with symmetric Rydberg states, the evolution of blockade radius is steady while being fluctuant for the asymmetric heteronuclear (or homonuclear) interactions. For Rb atom excited to Rydberg state $|77S_{\frac{1}{2}}, m_j=1/2\rangle$ interacting with Cs atom excited to Rydberg state $|81S_{\frac{1}{2}}, m_j=-1/2\rangle$, the value of blockade radius reaches a local maxima. Local maxima points are repeated also at $n=65$, and $122$ (the principal quantum corresponding to the Rydberg state of Cs atom). Local minima points ($n=67$, and $94$) can be imputed to principal quantum numbers where the dipole-dipole couplings of these pair states counteract each other. The evolution of Le Roy radius $R_{\text{\tiny LR}}$ is steady in both cases.

\section{Fidelity of multiqubit entangled states\label{sectionFidelity}}
Fidelity is a measure of the closeness of two arbitrary quantum states. We employed the standard definition \cite{nielsen2000quantum} of fidelity between arbitrary states $\hat{\rho}$ and $\hat{\sigma}$ of a quantum system as 
\begin{equation}
	F(\hat{\rho}, \hat{\sigma}) = \text{Tr} \left( \sqrt{ \sqrt{\hat{\rho}} \hat{\sigma} \sqrt{\hat{\rho}} } \right),
\end{equation}
where we considered $\hat{\rho}$ as the calculated density matrix after performing a partial trace over the subspace of computational states of control and target atoms, and $\hat{\sigma}=| \Phi^{+} \rangle \langle \Phi^{+} |$ is the density matrix of the multi-qubit entangled state  $|\Phi^{+}\rangle = \frac{1}{\sqrt{2}}\otimes_{i}^{k} \left(|0\rangle+ |1\rangle \right)_{i} \, \otimes_{j}^{N} \left(|A\rangle+ |B\rangle \right)_{j}$.
%\frac{1}{\sqrt{2}} \left( \otimes_{\ell}^{N} | 0 \rangle_{\ell} + \otimes_{\ell}^{N} | 1 \rangle_{\ell} \right)
We numerically calculated the density operator of the system when it was initially prepared in the superposition of  the computational ground states of control atom, which results in simultaneous  blocking and transferring operations $\frac{1}{\sqrt{2}}(|0\rangle|A^{N}\rangle+|1\rangle|A^{N}\rangle)\rightarrow\frac{1}{\sqrt{2}}(|0\rangle|A^{N}\rangle+|1\rangle|B^{N}\rangle)$ where $|A^{N}\rangle=\otimes_{k=1}^{N}|A\rangle_{k}$. The case when $N=1$ corresponds to a two-qubit  Bell state, while for $N>1$ we end in a GHZ-state, which is a useful resource in quantum computing and cryptography \cite{hillery1999quantum}. For $N=1$, we have  four computational states $|0 \, A\rangle, |0 \, B\rangle,|1 \, A\rangle$ and $|1 \, B\rangle$. Generally, the total number of basic computational states in Rydberg blockade of CNOT gate is equal to $2^{N+1}$ from the total $\mathcal{N}_{\text{\tiny C}}\times\mathcal{N}_{\text{\tiny T}}^{N}$ states, where $\mathcal{N}_{\text{\tiny C}\, (\text{\tiny T})}$ is the number of states in control (target) atom.

In our simulations, we consider different configurations of the control and target atoms to be either Rb ( $^{87}$Rb) or Cs ($^{133}$Cs). We have taken decay rates  $\gamma_{r}=1/\tau_c$ and $\gamma_{p}=1/\tau_p$ of the Rydberg state of the control atom $|r\rangle$ and the intermediate state of the target atoms $|P\rangle$, respectively, from the data of ARC \cite{vsibalic2017arc}.  The lifetime of Rb $|P\rangle=|5 P_{3/2}, m_j=3/2\rangle$ state (first excited state of Rb) is $\tau_{p}=26.4$~\si{n s} while for  Rb $|6 P_{3/2}, m_j=3/2\rangle$ state the lifetime is  $\tau_{p}=0.131\,\mu$s. The lifetime of Cs $|P\rangle= |6 P_{3/2}, m_j=3/2\rangle$ (first excited state of Cs) of the target atom is $\tau_p=30.5$~\si{n s}, while for Cs $|7P_{3/2}, m_j=3/2\rangle$ state we have $\tau_p=0.118$~$\mu\text{s}$. Since lifetimes of higher excited states of  target atoms are much longer, in the following calculations we consider Rb $|P\rangle=|6 P_{3/2}, m_j=3/2\rangle$  and  Cs $|7 P_{3/2}, m_j=3/2\rangle$ states as the intermediate state of the target atom. The Rydberg excitation schemes through these intermediate states were experimentally demonstrated in Refs. \cite{graham2022demonstration, levine2019parallel}. In section \ref{sectionGate-error}, we justify this choice of the intermediate state $|P\rangle$.

%----%----%----%----%----%----%----%----%----%----%----%----%----%----%----%----%----%----%----%----%----%----%----%----%----%----%----%----%----%----%----%----%----%----%----%----%----%----%----%----%----%----%----%----%----%----%----%----%----%----%----%----%----%----%----%----%----%----%----%----%----%----%----

\subsection{Homonuclear architecture}

\begin{figure}[htb!]\centering
	\subfloat[N=1]{
		\includegraphics[width=4cm,height=4cm]{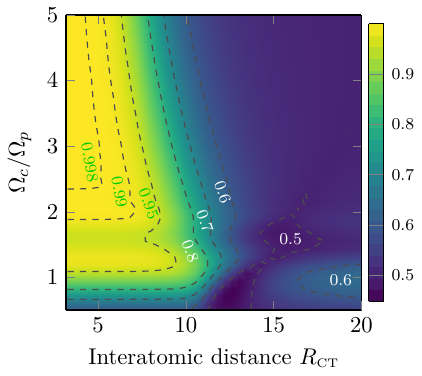}
	}
	\subfloat[N=2]{
		\includegraphics[width=4cm,height=4cm]{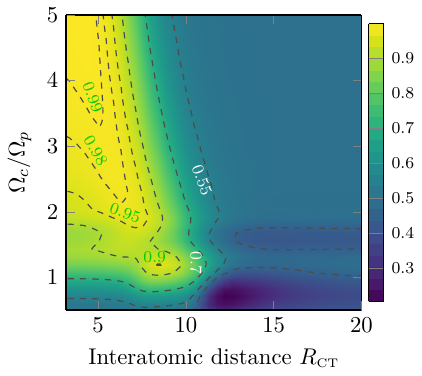}
	}\\
	\subfloat[N=3]{
		\includegraphics[width=4cm,height=4cm]{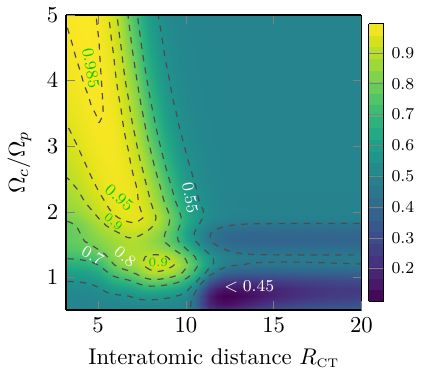}
	}
	\subfloat[N=4]{
		\includegraphics[width=4cm,height=4cm]{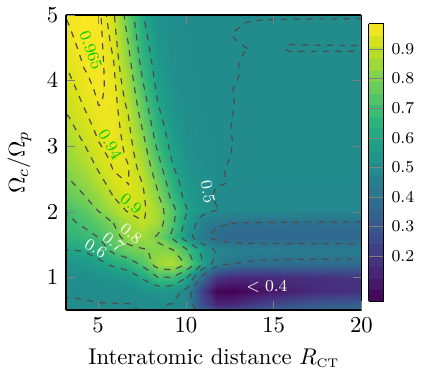}
	}
	\caption{Contour plot of the fidelity of entangled states $F_{\text{Rb-Rb}}$ as a function of interatomic distance $R_{\text{\tiny CT}}$~($\mu\text{m}$) and  coupling Rabi frequency  $\Omega_c/\Omega_p$, for homonuclear symmetric interaction of $|77 S_{1/2}, m_j=1/2\rangle$ Rb atoms for different spatial configurations of $N=1-4$ target atoms [see Figure~\ref{spatialconfiguration}].}
	\label{Fig: contours-RbRb-1}
\end{figure}

In this subsection, we study the gate performance for the case when control and target atoms are same atomic species, which can be either Rb or Cs. There are two different scenarios of homonuclear interaction: (i) symmetric interaction $\left|n S,n S\right\rangle \rightarrow \left| n' P , n'' P\right\rangle$, when both control and target atoms are excited to the same Rydberg state, and (ii) asymmetric interaction $\left|n S,\bar{n} S\right\rangle \rightarrow \left| n' P, \bar{n}' P\right\rangle$, when the control atom is excited to one Rydberg state with principal quantum number $n$, and the target atoms are excited to a different Rydberg state with principal quantum number $\bar{n}$. Note that in the latter case, the interaction between the target atoms is also symmetric (the quantum states $|\bar{n}S_{1/2}\rangle$ of target atoms are identical).  

\textit{--- Symmetric interaction channel $\left|n S,n S\right\rangle \rightarrow \left| n' P , n'' P\right\rangle$}. Figure~\ref{Fig: contours-RbRb-1} is a contour plot of fidelity of entangled states as a function of interatomic distance between control and target atoms $R_{\text{\tiny CT}}$~($\mu\text{m}$) and the ratio $\Omega_c/\Omega_p$ where we considered $|77 S_{1/2},m_j=1/2\rangle$ Rb  Rydberg states of control and target atoms with lifetime $\tau_c=505$~$\mu\text{s}$. In the case of only one target atom it is possible to achieve high fidelity $F=99.8\%$ for a wide range of interatomic distances $R_{\text{\tiny LR}}<R_{\text{\tiny CT}}<5.5$~$\mu\text{m}$ and moderate values of $\Omega_c \geq 2.5~ \Omega_p$, since the target-target interaction does not exist in this case. Considering schemes with more target atoms $N>1$, the fidelity for  $\Omega_c<2\,\Omega_p$ drops at small interatomic distances because of the increase of the influence of strong target-target interactions. The optimum interatomic distance is found to be around $R_{\text{\tiny CT}}\sim3-4$~$\mu\text{m}$. For $N=4$ target atoms, the fidelity $\approxeq 96.5\%$ for very high values of Rabi frequency $\Omega_c>3.5~\Omega_p$ at $R_{\text{\tiny CT}}=5~\mu\text{m}$ as in Figure~\ref{Fig: contours-RbRb-1}(d).

In principle, it is also possible to consider asymmetric homonuclear interactions ($\left|n S,m S\right\rangle \rightarrow \left| n' P , m' P\right\rangle$) to achieve high fidelities by reducing target-target interaction compared to control-target interaction. This case will be considered in a future work.

%----%----%----%----%----%----%----%----%----%----%----%----%----%----%----%----%----%----%----%----%----%----%----%----%----%----%----%----%----%----%----%----%----%----%----%----%----%----%----%----%----%----%----%----%----%----%----%----%----%----%----%----%----%----%----%----%----%----%----%----%----%----%----

\subsection{Heteronuclear architecture}

\begin{figure}[htb!]\centering
	\subfloat[N=1]{
		\includegraphics[width=4cm,height=4cm]{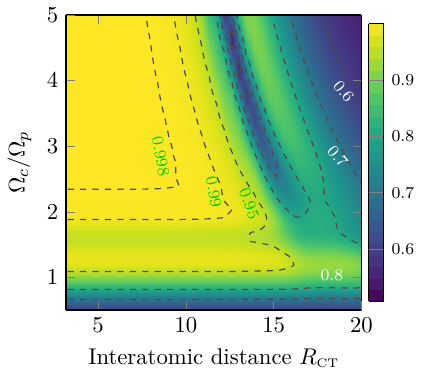}
	}
	\subfloat[]{
		\includegraphics[width=4cm,height=4cm]{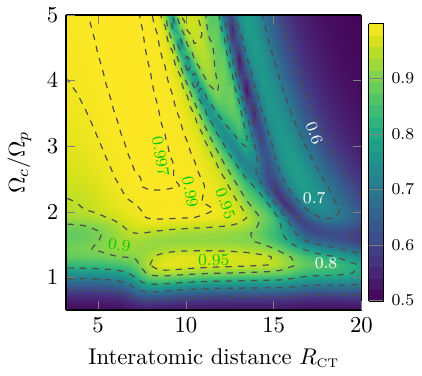}
	}\\
	\subfloat[N=3]{
		\includegraphics[width=4cm,height=4cm]{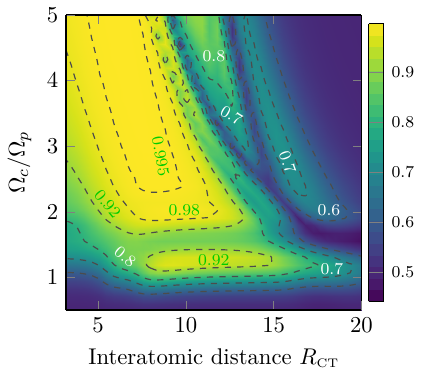}
	}
	\subfloat[N=4]{
		\includegraphics[width=4cm,height=4cm]{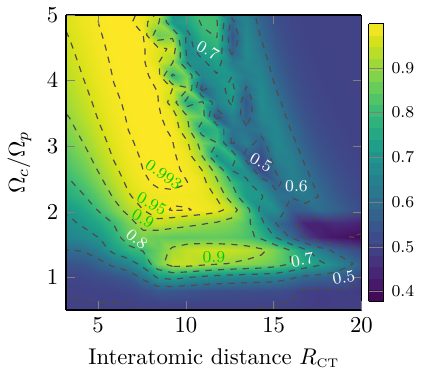}
	}
	\caption{Contour plot of fidelity $F_{\text{Cs-Rb}}$ as a function of interatomic distance $R_{\text{\tiny CT}}$~($\mu\text{m}$) and coupling Rabi frequency $\Omega_c/\Omega_p$, for the case of heteronuclear asymmetric interaction of Cs control atom  in state $|r\rangle=|81S_{\text{1/2}}, m_j=-1/2 \rangle$ and Rb target atoms in state $|R\rangle_{i}=|77S_{\text{1/2}}, m_j=1/2 \rangle$ for different spatial configurations of $N=1-4$ target atoms [see Figure~\ref{spatialconfiguration}].}
	\label{Fig: contours-CsRb}
\end{figure}

The contour plot of the fidelity  $F_{\text{\tiny Cs-Rb}}$ of entangled states in heteronuclear configuration is shown in Figure~\ref{Fig: contours-CsRb} as a function of interatomic distance $R$~($\mu\text{m}$) and the ratio $\Omega_c/\Omega_p$. This case corresponds to asymmetric heteronuclear interaction between control and target atoms, while the target atoms  interact in the vdW regime. The control Cs atom is excited to Rydberg state $|r\rangle=|81 S_{\text{1/2}}, m_j=-1/2 \rangle$ with lifetime $\tau_c=548$~$\mu\text{s}$ and Rb target atoms are excited to Rydberg state $|R\rangle_{i}=|77S_{\text{1/2}}, m_j=1/2 \rangle$. Heteronuclear configuration is clearly advantageous in terms of fidelity comparing to the symmetric homonuclear configuration, shown in Figure~\ref{Fig: contours-RbRb-1}. The regime of a CNOT gate with one target atom for the selected range of interatomic distances $R_{\text{\tiny LR}} \leq R_{\text{\tiny CT}}\leq10~\mu\text{m}$, as shown in Figure~\ref{Fig: contours-CsRb}(a), is governed purely by   dipole-dipole interaction and allows to achieve fidelity $F^{N=1}_{\text{Cs-Rb}} = 99.8 \%$ for $\Omega_c \geq 2.5~\Omega_p$. With increase of the number of target atoms $N>1$,  the maximum obtained fidelity slightly drops, similarly to the homonuclear case. The fidelity $F_{Cs-Rb}^{N=4} 99.3\%$ within the region $6~\mu\text{m} < R_{\text{\tiny CT}} < 10~\mu\text{m}$ and $\Omega_c > 2.5~\Omega_p$. This justifies the advantage of heteronuclear configuration for implementation of multiqubit CNOT$^{\text{N}}$ gates. Moreover, the two-species architecture is useful for improvement of readout without cross-talk when the state of a Rb data qubit is not affected by measurements performed by resonant light scattering by ancillary Cs atoms \cite{beterov2015rydberg}, which was recently demonstrated experimentally for arbitrary two-dimensional arrays of Rb and Cs atoms \cite{singh2022dual}.	

%----%----%----%----%----%----%----%----%----%----%----%----%----%----%----%----%----%----%----%----%----%----%----%----%----%----%----%----%----%----%----%----%----%----%----%----%----%----%----%----%----%----%----%----%----%----%----%----%----%----%----%----%----%----%----%----%----%----%----%----%----%----%----
\section{Scheme of Rydberg EIT C$_2$NOT$^2$\label{sectionC2NOT2}}
\begin{figure}[htb!] \centering
	\subfloat[]{
		\includegraphics[width=3cm,height=4cm]{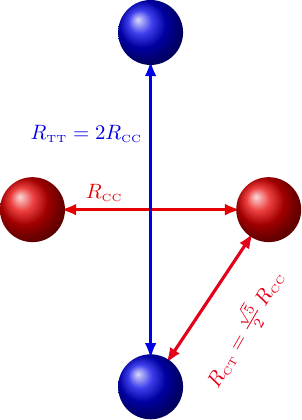}
	}
	\subfloat[]{
		\includegraphics[width=5cm,height=4cm]{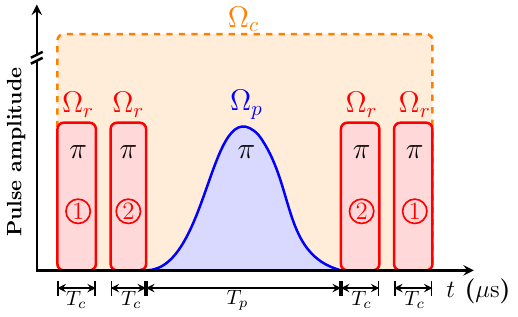}
	}	
	\caption{(a) Scheme of spatial configurations of homonuclear/heteronuclear interactions for the implementation of  C$_2$NOT$^2$ gate. The atoms are located on the vertices of a rhombus with perpendicular diagonals and $R_{\text{\tiny TT}}=2 \, R_{\text{\tiny CC}}$. $R_{\text{\tiny CC}}$ ($R_{\text{\tiny TT}}$) is the interatomic distance between control (target) atoms. (b) The sequence of laser pulses to perform C$_{2}$NOT$^{2}$ gate.}
	\label{modifiedpulsesequence}
\end{figure}

The proposals for implementation of  C$_{\text{k}}$NOT$^{\text{N}}$/C$_{\text{k}}$Z$^{\text{N}}$  gates with many control and many target atoms have been limited to  gates with either many control atoms and single target atoms,  or to single control atom and many target atoms. The most general case for arbitrary number of control and many target atoms has not been studied extensively. Recently, such schemes were proposed in several theoretical approaches \cite{young2021asymmetric,li2022multiple,cong2021hardware}. Young et al. \cite{young2021asymmetric} designed a protocol which uses microwave dressing to implement multi-qubit gates with  many control and many target atoms. This protocol reduces intraspecies interaction energies and maximizes the interspecies interaction energies, leading to asymmetric blockade, which simplifies the state preparation and enhances the speed of quantum algorithms and reduces the need for fault-tolerant error correction schemes. In this section, we modify the previously studied  CNOT scheme based on EIT in order to implement a four-qubit gate with two control and two target atoms (C$_2$NOT$^2$ gate) [see Figure~\ref{modifiedpulsesequence}(a)] simultaneously by proposing an asymmetric sequence of laser pulses acting on control atoms. We consider the following sequence [see the scheme in Figure~\ref{modifiedpulsesequence}(b)]: 
\begin{enumerate}
	\item We apply $\pi$-pulses to excite the control atoms from ground state $|1\rangle$ to highly excited Rydberg state $|r\rangle$ in sequence. 
\end{enumerate}
\begin{enumerate}\addtocounter{enumi}{1}
	\item Then we apply smooth Raman laser $\pi$ pulse to couple the ground states of the target atoms $|A\rangle$ and $|B\rangle$, simultaneously, to the intermediate dark state $|P\rangle$.
\end{enumerate}
\begin{enumerate}\addtocounter{enumi}{2}
	\item Finally, we apply $\pi$ pulses to return the control atoms from highly excited Rydberg state $|r\rangle$ to ground state $|1\rangle$ in reversed sequence applied in step 1.
\end{enumerate}

%----%----%----%----%----%----%----%----%----%----%----%----%----%----%----%----%----%----%----%----%----%----%----%----%----%----%----%----%----%----%----%----%----%----%----%----%----%----%----%----%----%----%----%----%----%----%----%----%----%----%----%----%----%----%----%----%----%----%----%----%----%----%----

%----%----%----%----%----%----%----%----%----%----%----%----%----%----%----%----%----%----%----%----%----%----%----%----%----%----%----%----%----%----%----%----%----%----%----%----%----%----%----%----%----%----%----%----%----%----%----%----%----%----%----%----%----%----%----%----%----%----%----%----%----%----%----

By applying this sequence of laser pulses, it is possible to efficiently implement the following gates by properly tuning the system parameters:
\begin{equation}\label{blockgate2-eq}
	\begin{split}
		\text{(I)No transfer: \hskip0.08cm}&|0\,0\rangle|AA\rangle \rightarrow 	|0\,0\rangle|AA\rangle, 
		\\& 
		| 0 \, 0 \rangle| BB \rangle \rightarrow 	| 0\,0 \rangle| BB \rangle,
		\\& 
		| 0 \,0 \rangle | A \,B \rangle \rightarrow 	|0\,0 \rangle | A \, B \rangle,
		\\& 
		| 0 \, 0 \rangle | B \, A \rangle \rightarrow  | 0 \, 0 \rangle |  B \, A \rangle,
	\end{split}
\end{equation}

%----%----%----%----%----%----%----%----%----%----%----%----%----%----%----%----%----%----%----%----%----%----%----%----%----%----%----%----%----%----%----%----%----%----%----%----%----%----%----%----%----%----%----%----%----%----%----%----%----%----%----%----%----%----%----%----%----%----%----%----%----%----%----

\begin{equation}\label{transfergate2-eq}
	\begin{split}
		\\ \text{(II)Transfer: }	&  |11\rangle | AA\rangle \rightarrow |11\rangle | BB\rangle, 
		\\&
		|11\rangle | BB\rangle \rightarrow |11\rangle | AA\rangle, 
		\\&
		|01\rangle | AB\rangle \rightarrow |01\rangle| BA\rangle, 
		\\&
		|01\rangle | BA \rangle \rightarrow |01\rangle | AB \rangle, 
		\\&
		| 10\rangle | AB \rangle \rightarrow | 10 \rangle | BA \rangle, 
		\\&
		|10\rangle | BA \rangle \rightarrow |10 \rangle| AB \rangle.
	\end{split}
\end{equation}

%----%----%----%----%----%----%----%----%----%----%----%----%----%----%----%----%----%----%----%----%----%----%----%----%----%----%----%----%----%----%----%----%----%----%----%----%----%----%----%----%----%----%----%----%----%----%----%----%----%----%----%----%----%----%----%----%----%----%----%----%----%----%----

\begin{figure}[htb!]\centering
	\subfloat[$V_{\text{\tiny CC}}\neq0$]{
		\includegraphics[width=4cm,height=4cm]{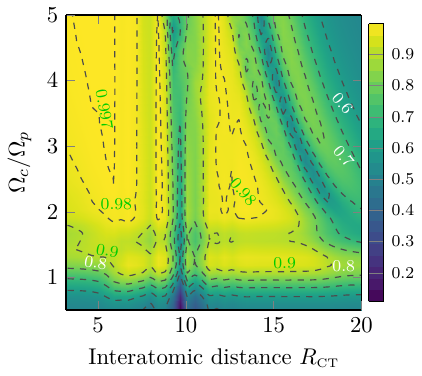}
	}
	\subfloat[$V_{\text{\tiny CC}}=0$]{
		\includegraphics[width=4cm,height=4cm]{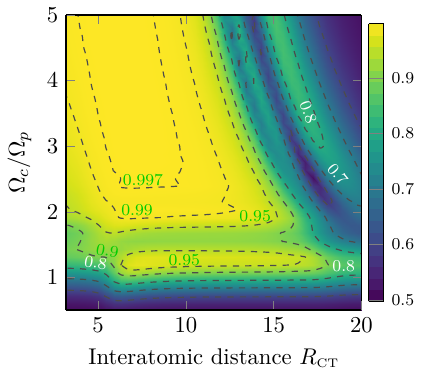}
	}\\
	\subfloat[]{
		\includegraphics[width=6cm,height=3.9cm]{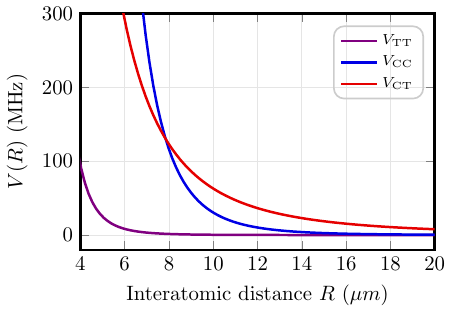}
	}
	\caption{
		Contour plot of fidelity $F_{\text{\tiny Cs-Rb}}^{C_2 NOT^2}$ for the case of heteronuclear interaction of Cs control atom  in state $|r\rangle_{i}=|81 S_{1/2}, m_j=-1/2 \rangle$ and Rb target atoms in state $|R\rangle_{j}=|77 S_{1/2}, m_j=1/2 \rangle$ as a function of the interatomic distance $R_{\text{\tiny CT}}$ and the ratio $\Omega_c/\Omega_p$. The system is initially prepared in state $\frac{1}{\sqrt{2}}\left( |00\rangle|AA\rangle + |11\rangle|AA\rangle \right)$. $\Omega_p=2\pi\times50$~\si{MHz}, and gate duration $\tau=1.53$~$\mu\text{s}$. (c) The evolution of interaction energies as a function of the interatomic distances according to the considered spatial arrangement in Figure \ref{modifiedpulsesequence}(a).}
	\label{Fig: contours-C2NOT2}
\end{figure}
The system dynamics for the case of only one control atom excited to the Rydberg state $|r\rangle$ and the second control atom remaining in the ground state $|g\rangle$ [i.e. $\frac{1}{\sqrt{2}}\big( |0\,0\rangle|A\,A\rangle + |0\,1\rangle|A\,A\rangle \big)\rightarrow \frac{1}{\sqrt{2}}\big( |0\,0\rangle |A\,A\rangle + |0\,1\rangle |B\,B\rangle \big)$ or $\frac{1}{\sqrt{2}}\big( |0\,0\rangle |A\,A\rangle + |1\,0\rangle |A\,A\rangle \big) \rightarrow \frac{1}{\sqrt{2}}\big( |0\,0\rangle |A\,A\rangle + |1\,0\rangle |B\,B\rangle \big)$] corresponds to the previously considered cases for homonuclear symmetric and for heteronuclear configurations in Figs.~\ref{Fig: contours-RbRb-1}(b), \ref{Fig: contours-CsRb}(b), respectively. 

The architecture of this gate becomes possible if the qubits satisfy the following conditions: (1) Cs control atoms are excited to Rydberg state $|81S_{1/2}, m_j = -1/2 \rangle$ where their dominant interaction regime is  vdW i.e. $V_{\text{\tiny CC}}=\frac{C_{6}}{R_{\text{\tiny CC}}^{6}}$, (2) Rb target atoms are excited to Rydberg state $|77S_{1/2}, m_j=1/2\rangle$ and similarly their dominant interaction regime is vdW i.e. $V_{\text{\tiny TT}}=\frac{C_{6}}{R_{\text{\tiny TT}}^{6}}$, (3)The regime of interaction between control Cs and target Rb atoms, is dipole-dipole interaction i.e. $V_{\text{\tiny CT}}=\frac{C_{3}}{R_{\text{\tiny CT}}^{3}}$.

In Figure~\ref{Fig: contours-C2NOT2}(a) and (b), we plot the contours of fidelity when the system is initially prepared in the superposition of ground states on control atoms, while the target atoms are in states $|A\, A\rangle$: $\frac{1}{\sqrt{2}}\left( |0\,0\rangle|A\,A\rangle + |1\,1\rangle|A\,A\rangle \right)\rightarrow \frac{1}{\sqrt{2}}\left( |0\,0\rangle|A\,A\rangle + |1\,1\rangle|B\,B\rangle \right)$ with gate duration of $\tau=8.06$~$\mu\text{s}$. The total number of computational basis states for a scheme with $k$ control atoms interacting with $N$ target atoms is equal to $2^{k+N}$.  In Figure~\ref{Fig: contours-C2NOT2}(a), we consider all possible interaction between control and target atoms. The maximum achieved fidelity is $99.7\%$ for a high value of Rabi frequency $\Omega_c/\Omega_p>2.5$ for $R_{\text{\tiny CT}}\approxeq6~\mu\text{m}$. It is also noticed a sharp drop in fidelity for $8~\mu\text{m}<R_{\text{\tiny CT}}<12~\mu\text{m}$ which can be a result of the interaction between control atoms, since in this regime interaction between target atoms almost vanishes as seen in Figure \ref{Fig: contours-C2NOT2}(c). In Figure~\ref{Fig: contours-C2NOT2}(b), we study an non-realistic case where we neglected the interaction between control atoms, which can be compared with the case of CNOT$^2$ in Figure \ref{Fig: contours-CsRb}(b), considering the different spatial arrangements. The maximum value of fidelity becomes possible for a wider range of intratomic distances. This case proves that the destructive pattern in system dynamics is a direct result of $V_{\text{\tiny CC}}$.

%----%----%----%----%----%----%----%----%----%----%----%----%----%----%----%----%----%----%----%----%----%----%----%----%----%----%----%----%----%----%----%----%----%----%----%----%----%----%----%----%----%----%----%----%----%----%----%----%----%----%----%----%----%----%----%----%----%----%----%----%----%----%----
\section{Gate errors \label{sectionGate-error}}

\begin{figure}[t]\centering
	\subfloat[CNOT$^{2}$]{
		\includegraphics[width=4cm,height=5cm]{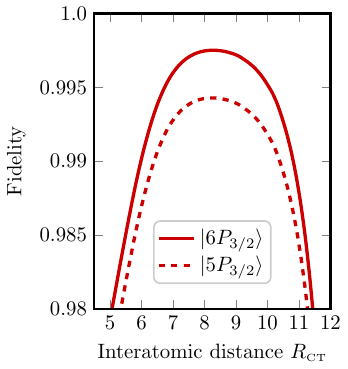}
	}
	\subfloat[CNOT$^{3}$]{
		\includegraphics[width=4cm,height=5cm]{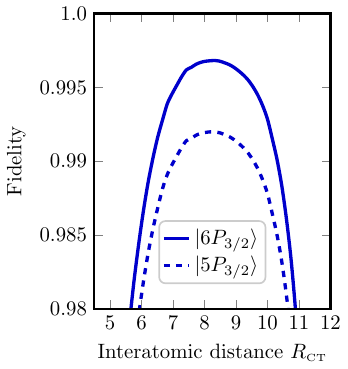}
	}\\
	\subfloat[C$_{2}$NOT$^{2}$]{
		\includegraphics[width=6cm,height=5cm]{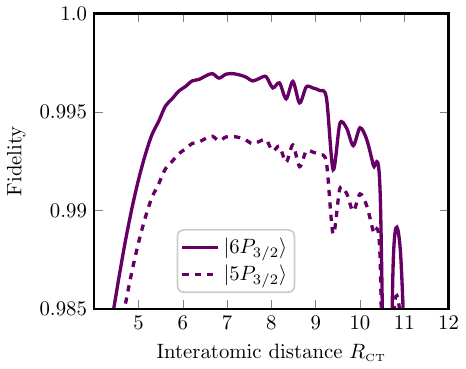}
	}
	\caption{Investigating the source of error resulting from exciting the target atom from ground state through the first of the second resonance level of intermediate state for different configurations: (a) CNOT$^2$, (b) CNOT$^{3}$, and (c) C$_{2}$NOT$^{2}$ gates. We plot the evolution of fidelity as a function of the interatomic distance $R_{\text{\tiny CT}}$ taking into account the finite lifetimes of excited states. Solid (dashed) curves represent the case of intermediate state of the Rb target atoms to be $|P\rangle=|6P_{3/2}, m_j=3/2\rangle$ ($|P\rangle=|5P_{3/2}, m_j=3/2\rangle$) [see main text.]. $\Omega_c=2.5~\Omega_p=2\pi\times 125$~\si{MHz}, $T_{\text{\tiny C}}=1$~$\mu\text{s}$.}
	\label{Fig: gate-error}
\end{figure}

In this section we discuss the effect of finite lifetimes, and the role of the excitation through different intermediate states and of the spatial arrangement of target atoms on the fidelity for of  CNOT$^{4}$ and C$_{2}$NOT$^{2}$ gates.

\textit{CNOT$^{\text{N}}$--- } In Figure~\ref{Fig: gate-error}(a) and (b), we plot the fidelity as a function of interatomic distance between Cs control and Rb target atoms for CNOT$^{N}$ gate for $N=2$, and $N=3$, respectively, with $\Omega_c=2.5~\Omega_p=2 \pi \times 125$~\si{MHz}, $T_{\text{\tiny C}}=1~\mu\text{s}$, and total gate time $\tau=3.28$~$\mu\text{s}$. Solid (dashed) curve represents the case of excitation of the target atoms through the second (first) resonance level of the intermediate state $|P\rangle=|6P_{3/2}\rangle$ ($|P\rangle=|5P_{3/2}\rangle$). It is clear that using the second resonance level enhanced the obtained fidelity for CNOT$^2$ (CNOT$^3$) to be $99.75\%$ ($99.68\%$) at $R_{\text{\tiny CT}}=8.33$~$\mu\text{m}$, compared with $99.43\%$ ($99.2\%$) for excitation through the first resonance level. %The blue curve represents the situation, where we omitted the effect of finite lifetimes of Rydberg state $|r\rangle$ of control atom and the intermediate state $|P\rangle$ of target atoms. In this case, the fidelity $F= 97.32 \%$ is limited by imperfect EIT conditions and population transfer through off-resonant interaction channels. 

\textit{C$_{k}$NOT$^{N}$--- } In Figure~\ref{Fig: gate-error}(c), we plot the fidelity of multi-control and multi-target C$_2$NOT$^{2}$ gate as a function of interatomic distance $R_{\text{\tiny CT}}$ between Cs control and Rb target atoms with total gate time $\tau=5.28$~$\mu\text{s}$. Solid (Dashed) curve  represents the case of excitation of the two target atoms through the second (first) resonance level of the intermediate state $|P\rangle=|6P_{3/2}\rangle$ ($|P\rangle=|5P_{3/2}\rangle$) which shows a possible fidelity $F=99.6\%$ ($99.3\%$) at $R_{\text{\tiny CT}}=8.4$~$\mu\text{m}$.

In our model, we considered the time gap between excitation and de-excitation of Rydberg state on the control atom to be $T_{p}$~($\mu$m) caused by the pulse sequence $\pi-\text{gap}-\pi$ which is typically required by a Rydberg blockade gate can cause an expected atom loss, not only for control atom but also for the ensemble of target atoms, which can also a source of errors for implementing the physical system in experiment. In \cite{li2023proposal}, the authors reported their findings  in constructing a native CNOT gate based on optimizing smooth Gaussian-shaped pulses.

%----%----%----%----%----%----%----%----%----%----%----%----%----%----%----%----%----%----%----%----%----%----%----%----%----%----%----%----%----%----%----%----%----%----%----%----%----%----%----%----%----%----%----%----%----%----%----%----%----%----%----%----%----%----%----%----%----%----%----%----%----%----%----

\begin{figure}[!htb]\centering
	\subfloat[$R_{\text{\tiny CT}}=5~\mu\text{m}$]{
		\includegraphics[width=8cm,height=6cm]{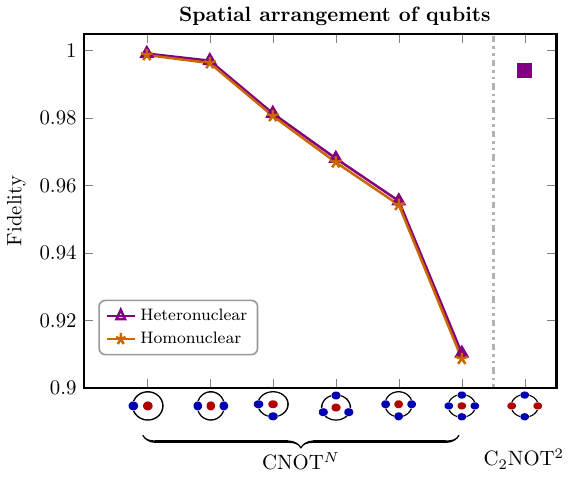}
	}\\
	\subfloat[$R_{\text{\tiny CT}}=7~\mu\text{m}$]{
		\includegraphics[width=8cm,height=6cm]{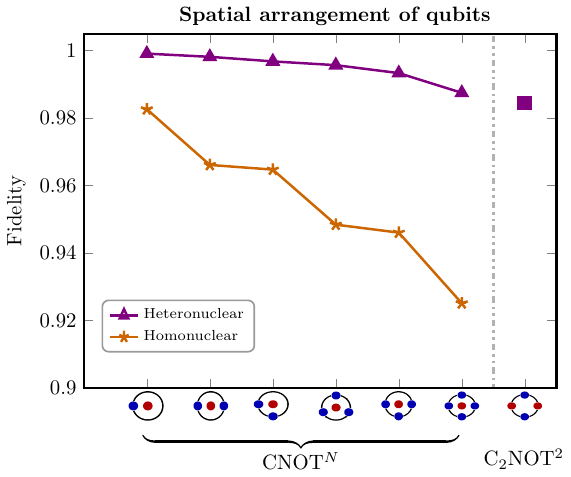}
	}
	\caption{The fidelity for different spatial arrangement of qubits in CNOT$^{N}$ (C$_{2}$NOT$^{2}$) gate for $\Omega_c \simeq 2.5~\Omega_p$ and interatomic distance $R_{\text{\tiny CT}} = 5$~$\mu\text{m}$ in (a), $R_{\text{\tiny CT}} = 7$~$\mu\text{m}$ in (b).}
	\label{Fig: summary}
\end{figure}

In Figure~\ref{Fig: summary}, we show the effect of spatial arrangement and the number of target atoms on the obtained fidelity of CNOT$^N$ gate for Rabi frequency $\Omega_c = 2.5\, \Omega_p = 2\pi\times 125$~\si{MHz} for two different values of $R_{\text{\tiny CT}}$. It is clear that the spatial arrangement of target atoms around the control atom slightly changes the fidelity according to their trapping positions. We also show that the heteronuclear architecture can be advantageous in terms of fidelity as a value for arrays with $R_{\text{\tiny CT}}>5~\mu\text{m}$.

%----%----%----%----%----%----%----%----%----%----%----%----%----%----%----%----%----%----%----%----%----%----%----%----%----%----%----%----%----%----%----%----%----%----%----%----%----%----%----%----%----%----%----%----%----%----%----%----%----%----%----%----%----%----%----%----%----%----%----%----%----%----%----

\section{Conclusion \label{conclusion}}

We studied the performance of the multi-qubit CNOT gates based on EIT and Rydberg blockade. Our simulations confirm the advantages of  heteronuclear architecture of the atomic quantum registers for suppression of the undesirable target-target interactions which limit the performance of multi-qubit gates. We have shown that in the configuration of single control and four target atoms which is most suitable for surface codes it is possible to achieve the fidelity of multi-qubit CNOT$^4$ gate above 99\% which opens the way to fast quantum error correction schemes with neutral atoms. 

\begin{acknowledgments}
We thank Mark Saffman for helpful discussions. This work is supported by the Russian Science Foundation (Grant No.~\href{https://rscf.ru/project/23-42-00031/}{23-42-00031}). A. Farouk is funded by the joint executive educational program between Egypt and Russia (EGY-6544/19). P. Xu acknowledges funding support from the National Key Research and Development Program of China (Grant \textnumero 2021YFA1402001), the Youth Innovation Promotion Association CAS \textnumero Y2021091. Datasets supporting the plots within this manuscript are available through Zenodo \cite{FaroukZenodoDataset2022}. Further information is available from the corresponding author. We also thank the anonymous referees for valuable comments.
\end{acknowledgments}

%----%----%----%----%----%----%----%----%----%----%----%----%----%----%----%----%----%----%----%----%----%----%----%----%----%----%----%----%----%----%----%----%----%----%----%----%----%----%----%----%----%----%----%----%----%----%----%----%----%----%----%----%----%----%----%----%----%----%----%----%----%----%----

\appendix

\section[\appendixname~\thesection]{Model of multiple Rydberg interaction channels}
\label{Appendix:Multi-Rydberg}
%\subsection[\appendixname~\thesubsection]{}

\begin{figure}[ht!]\centering
	\includegraphics[width=8cm,height=6.5cm]{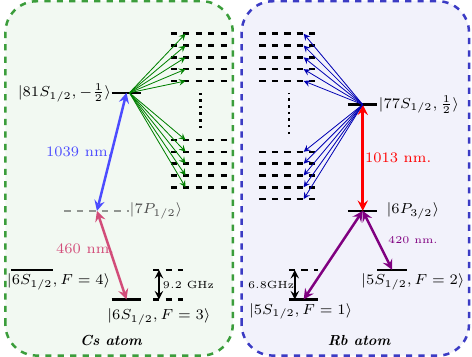}
	\caption{Hyperfine structure of the collective atomic energy levels with control $^{133}$Cs atom  and target $^{87}$Rb atom and laser-induced transitions between ground and Rydberg states.}
	\label{fig13:hyperfinestructure}
\end{figure}

In this Appendix, we consider that effect of multiple Rydberg interaction channels due to coupling of initially excited colective two-atom Rydberg state to numerous neighboring collective two-atom states through dipole-dipole interaction. This problem is well described in \cite{saffman2010quantum}, and its formalism is limitedly adopted in \cite{khazali2020fast, yu2022multiqubit}. In Ref.~\cite{yu2022multiqubit}, the authors showed that for implementation of Toffoli gate with homonuclear architecture using main interaction channel $|(n+1)\,S_{1/2} , n\,P_{3/2}\rangle \xtofrom[]{} |n\,P_{3/2},(n+1)\,S_{1/2}\rangle$ for dipole blockade the effect of other interaction channels (e.g. $|(n+1)\,S_{1/2} , n\,P_{3/2}\rangle \rightarrow |(n+2)\,P_{3/2},n\,S_{1/2}\rangle$) can be neglected. 

\begin{table*}
	\caption{The calculated dipole-dipole coefficient $\mathcal{C}_{3}^{(\alpha)}$ (\si{GHz}.$\mu\text{m}^3$), the energy defect $\delta_{E_{\alpha}}$ (\si{GHz}), the van der Waals $\mathcal{C}_6^{(\alpha)}$ (\si{GHz}.$\mu\text{m}^6$) interaction coefficients for: \textbf{Table (above)}, the heteronuclear asymmetric interaction channels $|r_0\rangle|R_0\rangle\rightarrow|r_\alpha\rangle|R_\alpha\rangle$ with $|r_0\rangle=|81S_{1/2},m_j=-1/2\rangle$ and $|R_0\rangle=|77S_{1/2},m_j=1/2\rangle$ for Cs and Rb atoms, respectively. \textbf{Table (below)}, the corresponding homonuclear symmetric interaction channels $|R_0\rangle |R_0\rangle \rightarrow |R_\alpha \rangle| R_\alpha \rangle $ for Rb atoms. $\omega_j$ ($j=1,2$) is the driving transition field that couples  $|r_0\rangle \mapsto |r_{\alpha}\rangle$ ($|R_0\rangle \mapsto |R_{\alpha}\rangle$). $\chi_{\alpha} = \dfrac{\mathcal{C}_{3}^{(\alpha)}}{ \bar{R}_{\text{\tiny CT}}^{3} \, \delta_{F_{\alpha}}}$ is the dimensionless coupling strength factor. If $\chi_{\alpha}\ll 1$, then the leakage from the resonantly coupled state can be suppressed. We limited the results of heteronuclear interaction to be $\Delta n=2$ and $\delta_{F_{\alpha}}/2\pi \, \in [-2,2]$~\si{GHz} and considered the corresponding interaction channels for the homonuclear interactions. $\bar{R}_{\textrm{\tiny CT}}=5$~$\mu\text{m}$, $\bar{R}_{\text{\tiny TT}}=\sqrt{2} \, \bar{R}_{\text{\tiny CT}}$. The values are taken from ARC using functions \texttt{getDipoleMatrixElement}, and \texttt{getEnergy}.}
	%\begin{ruledtabular}
	\begin{center}
		%	\begin{ruledtabular}
			\normalfont\addtolength{\tabcolsep}{5pt}
			\renewcommand{\arraystretch}{1.2}
			\begin{tabular}{||c||c|c|c|c|c||c|c|c|c|c||c|c|c|c||} \hline\hline
				\multirow{2}{*}{$\kappa$} & \multirow{2}{*}{$\omega_1$}& \multicolumn{4}{c||}{$^{87}$\textbf{Rb}} & \multirow{2}{*}{$\omega_2$} & \multicolumn{4}{c||}{$^{133}$\textbf{Cs}}& \multirow{2}{*}{$\mathcal{C}_3/2\pi$} & \multirow{2}{*}{$\delta_{E} / 2\pi$} & \multirow{2}{*}{$\chi$} & \multirow{2}{*}{$\mathcal{C}_6/2\pi$} \\ \cline{3-6} \cline{8-11}	%=========================================================
				& & $n_1$ & $\ell_1$ & $j_1$ & $m_{j1}$ & & $n_2$ & $\ell_2$ & $j_2$ & $m_{j2}$ & &  & & \\ \hline 	%=========================================================
				$1$ & $\sigma^+$ & $76$ & $1$ & $1.5$ & $1.5$ & $\sigma ^+$ & $81$ & $1$ & $0.5$ & $0.5$ & $10.9$ & $1.869$ & $4.650\times 10^{-2}$ & $5.812$ \\\hline
				%=========================================================
				$2$ & $\pi$  & $77$ & $1$ & $0.5$ & $0.5$ & $\sigma^+$ & $80$ & $1$ & $0.5$ & $0.5$ & $5.88$ & $0.2137$ & $2.201\times 10^{-1}$ & $27.52$ \\\hline
				%=========================================================
				$3$ & $\sigma^+$ & $77$ & $1$ & $1.5$ & $1.5$ & $\sigma ^+$ & $80$ & $1$ & $0.5$ & $0.5$ & $10.0$ & $0.0022$ & $3.656\times 10^{+1}$ & $4570$ \\\hline
				%=========================================================
				$4$ & $\pi$  & $78$ & $1$ & $0.5$ & $0.5$ & $\sigma^+$ & $79$ & $1$ & $0.5$ & $0.5$ & $0.105$ & $-0.4367$ & $1.927\times 10^{-3}$ & $-0.2409$ \\\hline
				%=========================================================
				$5$ & $\sigma^+$ & $78$ & $1$ & $1.5$ & $1.5$ & $\sigma^+$ & $79$ & $1$ & $0.5$ & $0.5$ & $0.190$ & $-0.6398$ & $2.374\times 10^{-3}$ & $-0.2968$ \\ \hline\hline
				%=========================================================
			\end{tabular}
			\vskip0.5cm
			\begin{tabular}{||c||c|c|c|c|c||c|c|c|c|c||c|c|c|c||} \hline\hline
				\multirow{2}{*}{$\kappa$} & \multirow{2}{*}{$\omega_1$}& \multicolumn{4}{c||}{$^{87}$\textbf{Rb}} & \multirow{2}{*}{$\omega_2$} & \multicolumn{4}{c||}{$^{87}$\textbf{Rb}}& \multirow{2}{*}{$\mathcal{C}_3/2\pi$} & \multirow{2}{*}{$\delta_{E} / 2\pi$} & \multirow{2}{*}{$\chi$} & \multirow{2}{*}{$\mathcal{C}_6/2\pi$} \\ \cline{3-6} \cline{8-11}	%=========================================================
				& & $n_1$ & $\ell_1$ & $j_1$ & $m_{j1}$ & & $n_2$ & $\ell_2$ & $j_2$ & $m_{j2}$ & &  & & \\ \hline 	%=========================================================
				$1$ & $\sigma^+$ & $76$ & $1$ & $1.5$ & $1.5$ & $\sigma^+$ & $76$ & $1$ & $1.5$ & $1.5$ & $11.2$ & $16.84$ & $6.657 \times 10^{-4}$ & $67.26$  \\ \hline
				%=========================================================
				$2$ & $\pi$  & $77$ & $1$ & $0.5$ & $0.5$ & $\pi$  & $77$ & $1$ & $0.5$ & $0.5$ & $4.29$ & $-15.40$ & $2.784 \times 10^{-4}$ & $-96.78$  \\ \hline
				%=========================================================
				$3$ & $\sigma^+$ & $77$ & $1$ & $1.5$ & $1.5$ & $\sigma^+$ & $77$ & $1$ & $1.5$ & $1.5$ & $12.5$ & $-15.82$ & $7.873 \times 10^{-4}$ & $-88.39$ \\ \hline
				%=========================================================
				$4$ & $\pi$  & $78$ & $1$ & $0.5$ & $0.5$ & $\pi$  & $78$ & $1$ & $0.5$ & $0.5$ & $0.0605$ & $-46.79$ & $1.294 \times 10^{-6}$ & $-0.00635$ \\ \hline
				%=========================================================
				$5$ & $\sigma^+$ & $78$ & $1$ & $1.5$ & $1.5$ & $\sigma^+$ & $78$ & $1$ & $1.5$ & $1.5$ & $0.197$ & $-47.20$ & $4.178 \times 10^{-6}$ & $-0.00743$ \\ \hline\hline
				%=========================================================
			\end{tabular}
			\label{Table:NewRydbergHeteroC3Table}
			%	\end{ruledtabular}
	\end{center}
	%\end{ruledtabular}
\end{table*}

%----%----%----%----%----%----%----%----%----%----%----%----%----%----%----%----%----%----%----%----%----%----%----%----%----%----%----%----%----%----%----%----%----%----%----%----%----%----%----%----%----%----%----%----%----%----%----%----%----%----%----%----%----%----%----%----
%----%----%----%----%----%----%----%----%----%----%----%----%----%----%----%----%----%----%----%----%----%----%----%----%----%----%----%----%----%----%----%----%----%----%----%----%----%----%----%----%----%----%----%----%----%----%----%----%----%----%----%----%----%----%----%----%----%----%----%----%----%----%----

\begin{figure}[!thb]\centering
	\subfloat[]{
		\includegraphics[width=4cm,height=4cm]{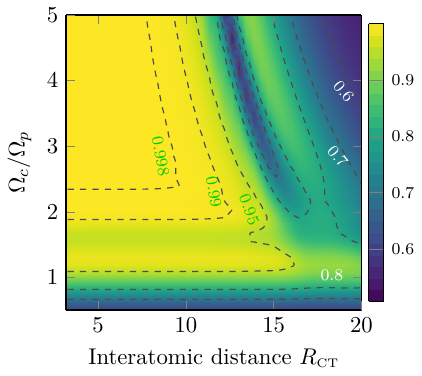}
	}
	\subfloat[]{
		\includegraphics[width=4cm,height=4cm]{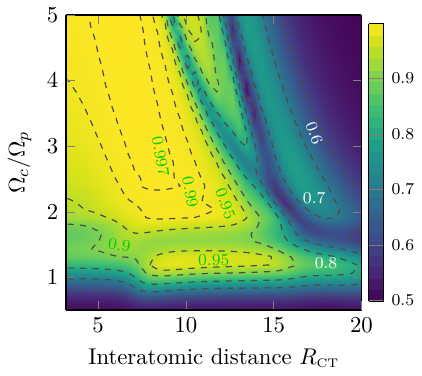}
	}
	\caption{(a) The contour plots of fidelity for implementing CNOT$^{N}$ gate using multi-Rydberg model as a function of $R_{\textrm{\tiny CT}}$ and $\Omega_c/\Omega_p$ for (a) $N=1$, (b) $N=2$.}
	\label{fig14:MultiRydbergCNOT}
\end{figure}

%----%----%----%----%----%----%----%----%----%----%----%----%----%----%----%----%----%----%----%----%----%----%----%----%----%----%----%----%----%----%----%----%----%----%----%----%----%----%----%----%----%----%----%----%----%----%----%----%----%----%----%----%----%----%----%----%----%----%----%----%----%----%----%----%----%----%----%----%----%----%----%----%----%----%----%----%----%----%----%----%----%----%----%----%----%----%----%----%----%----%----%----%----%----%----%----%----%----%----%----%----%----%----%----%----%----%----%----%----%----%----%----%----%----%----%----%----%----%----%----%----%----%----%----%----%----%----%----%----%----%----%----%----%----%----%----%----%----%----%----%----%----%----%----%----%----%----%----%----%----%----%----%----%----%----%----%----%----%----%----%----%----%----%----%----%----%----%----%----%----%----%----%----%----%----%----%----%----%----%----%----%----%----%----%----%----%----%----%----

As in Figure~\ref{fig13:hyperfinestructure}, we will consider the state $| r_0;R_0\rangle = |81S_{1/2},m_j=-1/2 \, ; \, 77S_{1/2},m_j=1/2 \rangle $ which is coupled to a set of other Rydberg states $\{|r_1,R_1\rangle ,|r_2,R_2\rangle, \dots \}$ with $|r_{\alpha}\rangle=|(81 \pm \bar{n} ) P_{j}, m_j \rangle$ and $|R_{\alpha}\rangle=|(77\pm \bar{m} ) P_{j}, m_j \rangle$ for different values of total angular momentum $j$ and of the projection of the total angular momentum on the quantization $z$-axis $m_j$, as we show in Table \ref{Table:NewRydbergHeteroC3Table}, where $\Delta n$ is the range of variation of principal quantum number $n$. $\bar{n}$ and $\bar{m}$ are integers  $\in \, [-\Delta n, \Delta n]$. We limited our calculations here for  $\Delta\ell=1$ for the interaction channel $|ss\rangle \rightarrow |pp\rangle$.  The control-target interaction Hamiltonian in equation (\ref{Control-TargetinteractionHamiltonian}), can be straightforwardly  modified to include coupling to many other Rydberg states.

Energy defect $\delta_{F_{\alpha}}$ is the energy difference between any two states for the interaction channel $\alpha$,
\begin{equation}
	\begin{split}
		\delta_{F_{\alpha}}&=\delta_{|r_0\rangle \rightarrow |r_{\alpha}\rangle}+\delta_{|R_0\rangle \rightarrow |R_{\alpha}\rangle},
		\\&=\big(E_{|r_0\rangle} - E_{|r_{\alpha}\rangle}\big) + \big(E_{|R_0\rangle} - E_{|R_{\alpha}\rangle}\big).
	\end{split}\label{EnergyDefect}
\end{equation}
where $E_{|\lambda\rangle}$ is the energy of atomic state $| \lambda \rangle$ with respect to the center of gravity of the hyperfine-split states.

The dipole-dipole $\mathcal{C}_3^{(\alpha)}$ coefficient can be calculated for specific interaction channel $\alpha$ by considering the dipole matrix element $\langle n_{0} , \ell_{0} , j_{0} , m_{j_{0}}| \, e \,\textbf{r} \, | n_{\alpha} , \ell_{\alpha} , j_{\alpha} , m_{j_{\alpha}}\rangle$. Adopting the notation $|\lambda_0\rangle=| n_{0} , \ell_{0} , j_{0} , m_{j_{0}}\rangle$ and $|\lambda_{\alpha}\rangle=| n_{\alpha} , \ell_{\alpha} , j_{\alpha} , m_{j_{\kappa}} \rangle$ for the initial and final Rydberg states, respectively. $\mathcal{C}_{3}^{(\alpha)}$ and the van der Waals coefficient $\mathcal{C}_{6}^{(\alpha)}$ can be calculated from the following forms ($\epsilon_{0}$ is the permittivity of free space):
\begin{equation}
	\begin{split}
		\mathcal{C}_{3}^{(\alpha)}=\frac{1}{4 \pi\epsilon_{0}} \langle r_0 | \, e \, \textbf{r} \, |r_{\alpha} \rangle \langle R_0 | \, e \, \textbf{r} \, |R_{\alpha} \rangle ,
	\end{split}\label{NEWdispersiveC3}
\end{equation}

\begin{equation}
	\begin{split}
		\mathcal{C}_{6}^{(\alpha)}=\frac{1}{4 \pi \epsilon_{0}}\frac{| \langle r_0 | \, e \, \textbf{r} \, |r_{\alpha} \rangle |^2 \, | \langle R_0 | \, e \, \textbf{r} \, |R_{\alpha} \rangle |^2}{ \delta_{F_{\alpha}} },
	\end{split}\label{NEWdispersiveC6}
\end{equation}
with a cross-over between these two-regimes occurs at radius $R_{\text{\tiny vdW}}=\left( \, C_{3}^{(\alpha)\,2}/\delta_{F_{\alpha}}^{2} \, \right)^{1/6}$. In Table \ref{Table:NewRydbergHeteroC3Table}, we show the values of $\mathcal{C}_3^{(\alpha)}$, $\mathcal{C}_6^{(\alpha)}$ and $\delta_{F_{\alpha}}$ for different interaction channels.

For the interaction with many target atoms, the target-target interaction Hamiltonian in equation (\ref{Target-TargetinteractionHamiltonian}) can be written to include more Rydberg states.

The values of $\mathcal{C}_{3}^{(\alpha)}$ and $\delta_{F_{\alpha}}$ for the symmetric homonuclear interaction between Rb atoms are given in Table \ref{Table:NewRydbergHeteroC3Table}[below]. The dimensions of the system $\mathbb{D}_N$ with many Rydberg states substantially increases. The system dimensions are written as $\mathbb{D}_N = \left( \mathcal{N}_{\text{\tiny C}} + \mathcal{K} \right) \times \left( \mathcal{N}_{\text{\tiny T}} + \mathcal{K} \right)^{N}$, where $\mathcal{K}$ is the number of considered other coupled Rydberg states for $N$ of target atoms.

In Figure~\ref{fig14:MultiRydbergCNOT}, we show the contour plots of fidelity as a function of interatomic distance $R_{\textrm{\tiny CT}}$ and $\Omega_c/\Omega_p$ using the model of coupling of the initially excited Rydberg state to many other Rydberg states for $N=1$ and $2$. It is clearly seen that the coupling to other Rydberg states does not result in significant changes to the value of the calculated fidelity compared to their results shown in Figures \ref{Fig: contours-CsRb}(a) and (b), where only single interaction channel was considered. The difference in fidelities between these models is of order $10^{-3}\sim10^{-7}$ for $N=1$, and of order $10^{-3}\sim10^{-5}$ for $N=2$.

For greater number of target atoms with using many Rydberg states results in substantial increase of the dimension of the model, which makes calculations difficult. Therefore we limited to a single-channel model in our calculations. The situation it the same for many control atoms.  Dongmin et al. \cite{yu2022multiqubit}, demonstrated that non-resonant couplings between the identical Rydberg states for control atoms in Toffoli gates can be regarded as weak leakage error. We checked our model for a wider range of $\Delta n$ and higher values $\delta_{F_{\alpha}}$ for the heteronuclear control-target interaction, and got the same conclusion. 

%%%%%%%%%%%%%%%%%%%%%%%%%%%%%%%%%%%%%%%%%%%%%%%%%%%%%%%%%%%%%%%%%%%%%%%%%%%%%%%%%%%%%%%%%%%%%%%%%%%%%%%%%%%%%%%%%%%%%%%%%%%%%%%%%%%%%%%%%%%%%%%%%%%%%%%%%%%%%%%%%%%%%%%%%%%%%%%%%%%%%%%%%%%%%%%%%%%%%%%%%%%%%%%%%%%%%%%%%%%%%%%%%%%%%%%%%%%%%%%%%%%%%%%%%%%%%%%%%%%%%%%%%%%%%%%%%%%%%%%%%%%%%%%%%%%%%%%%%%%%%%%%%%%%%%%%%%%%%%%%%%%%%%%%%%%%%%%%%%%%%%%

\nocite{*}
\bibliography{apssamp,apstemplateNotes}% Produces the bibliography via BibTeX.

%----%----%----%----%----%----%----%----%----%----%----%----%----%----%----%----%----%----%----%----%----%----%----%----%----%----%----%----%----%----%----%----%----%----%----%----%----%----%----%----%----%----%----%----%----%----%----%----%----%----%----%----%----%----%----%----%----%----%----%----%----%----%----

\end{document}